\documentclass{PoS}
\usepackage[authoryear,square]{natbib}
\bibpunct{(}{)}{;}{a}{}{,}
\usepackage{myitemize}
\usepackage{xspace}
\usepackage[pdftex]{graphicx}

\usepackage{multicol}
\usepackage{rotating} 
\usepackage{array}
\newcolumntype{L}[1]{>{\raggedright\let\newline\\\arraybackslash\hspace{0pt}}m{#1}}
\newcolumntype{C}[1]{>{\centering\let\newline\\\arraybackslash\hspace{0pt}}m{#1}}
\newcolumntype{R}[1]{>{\raggedleft\let\newline\\\arraybackslash\hspace{0pt}}m{#1}}

\newcommand{\specialcell}[2][c]{%
  \begin{tabular}[#1]{@{}l@{}}#2\end{tabular}}

\let\ifpdf\relax

\title{Connecting the Baryons: Multiwavelength Data for SKA HI Surveys}

\ShortTitle{Multiwavelength Data for SKA HI Surveys}

\author{\speaker{Martin Meyer}, Aaron Robotham, Danail Obreschkow, Simon Driver, Lister Staveley-Smith\\
International Centre for Radio Astronomy Research (ICRAR)\\
        E-mail: \email{martin.meyer@uwa.edu.au}}
\author{Martin Zwaan\\
European Southern Observatory\\}

\abstract{
The science achievable with SKA \hi surveys will be greatly increased through the combination of \hi data with that at other wavelengths. These multiwavelength datasets will enable studies to move beyond an understanding of \hi gas in isolation to instead understand \hi as an integral part of the highly complex baryonic processes that drive galaxy evolution.\\

As they evolve, galaxies experience a host of environmental and feedback influences, many of which can radically impact their gas content. Important processes include: accretion (hot and cold mode, mergers), depletion (star formation, galactic winds, AGN), phase changes (ionised/atomic/molecular), and environmental effects (ram pressure stripping, tidal effects, strangulation). Governing all of these to various extents is the underlying dark matter distribution. In turn, the result of these processes can significantly alter the baryonic states in which material is finally observed (stellar populations, dust, chemistry) and its morphology (galaxy type, bulge/disk ratio, bars, warps, radial profile). To fully understand the evolution of \hi and the role it plays in galactic evolution requires the ability to quantify each of these separate processes, and hence to coordinate SKA \hi surveys with extensive multi-band photometric and spectroscopic campaigns. In addition, multiwavelength data is essential for statistical methods of \hi analysis such as \hi stacking and intensity mapping cross-correlations.\\

In this chapter, we examine some of the principal science motivations for acquiring multiwavelength data to match that from the extragalactic SKA \hi surveys, and review the currently planned capacity to achieve this (eg. LSST, Euclid, W-FIRST, SPICA, ALMA, and 4MOST).  }

\FullConference{
Advancing Astrophysics with the Square Kilometre Array\\
June 8-13, 2014\\
Giardini Naxos, Italy}

\newcommand{\skipthis}[1]{}

\newcommand{\hi}{{\sc Hi}\xspace}
\newcommand{\ohi}{$\Omega_{\rm HI}$\xspace}

\newcommand{\be}{\begin{equation}}
\newcommand{\ee}{\end{equation}}

\begin{document}

\section{Introduction}

Our knowledge of galaxies and their evolution has been dominated by studies of the late-stage products of the evolutionary cycle, with extensive surveys in the optical and nearby wavelengths tracing the evolution of stellar material across large fractions of cosmic time.  In contrast, our knowledge of the \hi content of galaxies, the fundamental baryonic material out of which galaxies are made and a key tracer of their dynamics, remains rudimentary. Direct measurements of \hi content in individual galaxies beyond the local universe are currently limited to galaxy samples of only hundreds out to z$\sim$0.2-0.3 (BUDHIES, \citealt{Verheijen:2007p4033}; CHILES, \citealt{fernandez2013}), compared to the multi-million object samples out to z$\sim$1 and beyond that will be possible with the SKA (more than half the history of the Universe).  Linking these groundbreaking \hi samples with matched multiwavelength tracers will provide transformational new datasets spanning all of the major galactic constituents, and provide unique insight into the evolution of galaxies across cosmic time.

In this chapter, we explore the needs and availability of multiwavelength data for extragalactic \hi surveys with the SKA. Section~\ref{sec:need} summarises some of the principal scientific drivers for ancillary multiwavelength data and the analysis techniques these enable, followed by an overview of upcoming multiwavelength survey facilities in Section~\ref{sec:facilities}.  The implications for future SKA and multiwavelength survey design are discussed in Section~\ref{sec:design}.  Finally, conclusions are given in Section~\ref{sec:conclusion}, including a basic ranking of facilities and multiwavelength science products based on the degree to which well-matched multiwavelength data will be possible for SKA \hi surveys, as known at the time of writing. 

\vspace{-2mm}
\section{Need to Match HI Surveys with Multiwavelength Data}
\label{sec:need}
\vspace{-2mm}

\subsection{Characterising Different Baryonic States}

Understanding the mass assembly of galaxies, and the evolution of material from one baryonic state to another, is at the heart of galaxy evolution studies.  A large range of processes and scales are involved in this assembly, from the on-going gravitational collapse of gas into the filamentary structures of the cosmic web and its accretion into galactic halos, the cooling of gas into galactic disks and dense molecular clouds, to the subsequent star formation processes within these structures.  In this simple picture, galactic \hi acts as a bridge between the reservoir of gas in the cosmic web and the stellar material in galaxies.  However, the full picture of \hi content and its role in galaxy evolution is likely much more complex, with a host of mechanisms that can variously act to both add and remove \hi from galaxies.  Potentially important processes here include: accretion (hot and cold mode, mergers; eg. \citealt{Keres:2005p3529}), depletion (star formation, galactic winds, AGN; eg. \citealt{power2010,kim2013}), phase changes (ionised/atomic/molecular; eg. \citealt{hopkins2008}), and environmental effects (eg. \citealt{gunn1972,moore1996}).  Each of these sources and sinks ideally needs to be quantified through multiwavelength observations to determine the relative importance of each, and the nature of the fundamental scaling relations that link them (eg. the Schmidt law, \citealt{schmidt1959}; Table~\ref{fig:tracers} provides a summary of some of the central multiwavelength tracers and the physical processes and galactic components they measure).  The SKA will provide an unprecedented window through which to study the role of \hi in these relations across a large fraction of the history of the Universe, and will do so on both a global and morphologically resolved basis.  

\subsection{Characterising Galaxy Dynamics}

Along with mass, another key property in the evolution of galaxies are their kinematics and angular momenta. \hi observations with the SKA will be a powerful addition to this field, with SKA1 alone expected to yield two orders of magnitude more good \hi kinematic maps than the leading existing datasets \citep{Obreschkow:2014vd}, and more than a million galaxies with spatially unresolved \hi profiles.  While a powerful diagnostic in their own right, with \hi gas extending further into the dark matter haloes of galaxies than can be typically traced by stellar emission, the potential of these data will be best exploited through its combination with tracers at other wavelengths.  At the most basic level, higher resolution optical photometry enables inclination corrected rotation velocities to be extracted from line-of-sight global \hi profiles, along with improved morphological parameters such as disk scale lengths.  At a more detailed level, matched optical/NIR spectroscopy and IFU data provide complimentary measurements of the stellar-phase and ionized gas-phase material, which combined with \hi data will allow the full kinematic and angular momentum properties of galaxies to be studied, linking the often different evolutionary information encoded in the kinematics of gas and stars, and indeed tracing the dynamics of different morphological features such as disks and bulges.  Through both joint and separate analyses, \hi and multiwavelength data will enable the full suite of galactic dynamical scaling relations to be examined (eg. \citealt{tully1977}; \citealt{faber1976}; M$_{*}$-$S_{0.5}$ relation, \citealt{kassin2007}), and their evolution, underlying physics, and observational biases to be understood.

\begin{table}[t]
\begin{center}
\small
\begin{tabular}{p{4cm}p{10cm}}
\hline
Property                       &   Tracers   \\
\hline
atomic gas mass        & \hi \\
molecular gas mass  & mm emission lines (eg. CO) \\
stellar mass                 &  optical/NIR multiband photometry (eg. K, g-i), optical spectroscopy  \\
dust mass/temperature            & mid/far-IR multiband photometry, (sub)mm continuum \\
morphology                 &  panchromatic high resolution imaging, fitted structural parameters and decomposition (bulge, disk, bar)\\
star formation rate      &  H$\alpha$, UV, radio continuum, mid-IR\\
supermassive black hole     &   optical emission line diagnostics, radio continuum\\
chemistry                      &  optical/submm/mm spectroscopy via emission/absorption lines (high R)\\
dynamics                     & line fitting (optical, near-IR, CO, HI), spatially resolved (optical IFU, CO, HI) \\
environment               & redshift surveys (low R) + group catalogues (membership, multiplicity, halo mass, central/satellite classification), photo-z (multiband optical/NIR, large scale structure), X-rays (clusters, hot groups, massive halos), radio SZ (clusters, high-z)\\
distance                       & redshift (Hubble flow), Tully-Fisher (HI+optical/NIR imaging), Faber Jackson/Fundamental Plane (optical/NIR imaging and spectroscopy)\\
\hline
\end{tabular}
\end{center}
\caption{Physical quantities and their multiwavelength tracers}
\label{fig:tracers}
\end{table}%

\subsection{Characterising Environment}

Some of the most important external effects driving the distribution of galaxy properties we observe today are the environmental factors that can drastically impact \hi content.  For instance, the suppression of star formation in cluster environments is not due to an alteration in the actual processes of star formation, but rather due to a quenching of the \hi supply (tidal influences, \citealt{moore1996}; ram pressure stripping, \citealt{gunn1972}; strangulation, \citealt{balogh2000}).  However, the detailed processes by which environment affects \hi remain only partially understood and will be an important area of research for the SKA.  In the case of star formation quenching, for example, major optical surveys such as 2dFGRS find that star formation is suppressed at large distances from clusters, implying that evolutionary processes must be at work beyond these environments and the simple ram-pressure stripping of \hi gas in their cores \citep{lewis2002}.  Existing studies also find contradictory results on the environmental dependence of the \hi mass function \citep{zwaan2005,springob2005}. 

A potential contributor to some of the observational uncertainties and disagreements in existing environmental effect results are difficulties associated with properly characterising environment and the variety of different methods that have been used (eg. \hi vs. optical metrics). From an observational perspective, overcoming these challenges requires the combination of \hi data with deep, well-sampled optical redshift samples to provide a full characterisation of environment (group membership, group multiplicity, parent halo mass, central/satellite status). Together, these datasets will sample a greater dynamic range of environment than is possible from examining the distribution of \hi-rich galaxies alone, which are strongly anti-biased in the local Universe.  Direct tracers of the intra-group/cluster medium, such as that provided by X-ray data, will also be a highly valuable addition to the group catalogues provided by redshift surveys.  Interferometric \hi observations with sufficient resolution to isolate emission from galaxies in close proximity will also be needed to mitigate confusion effects in these environments, a requirement that will be met over a large redshift range by the SKA.

\subsection{Characterising Feedback}

The feedback effects from local processes such as star formation and AGN activity can be equally as important as those on larger scales.  For instance, recent simulation results find that feedback from supernovae impacts the overall normalisation of the \hi mass function, while AGN feedback can impact its high mass slope, with increased activity decreasing the number of high \hi mass galaxies observed \citep{kim2013}. The strength of these feedback mechanisms is also found to have an inverse effect on the global clustering strength of the \hi-rich galaxy population. Importantly, the various feedback mechanisms are also found to imprint differently on the distribution functions of the \hi-rich galaxy population compared to that observed at other wavelengths, offering an improved diagnostic ability for understanding the role of feedback in galaxy evolution when \hi datasets are combined with those selected, for instance, in the optical.  Nevertheless, a great deal of progress remains to be made in this area, and favoured models still offer sometimes widely discrepant results with observations \citep{duffy2012}.  Large-scale \hi surveys with the SKA, combined with matching multiwavelength data, will be essential for resolving these uncertainties, tracing the causes and varying effects of feedback across all of the major galactic constituents.

\subsection{Enabling Alternative Analysis Techniques}

Finally, multiwavelength data can also offer new ways of analysing \hi datasets.  Notably, the combination of \hi data with independent optical redshift catalogues has been used to vastly increase the redshift range over which measurements of the cosmic \hi density and gas fraction scaling dependencies can been made.  In one technique, \hi stacking, the rest frame \hi spectra of a large number of sources - extracted from an \hi datacube and shifted to rest frame on the basis of their known optical positions and redshifts - are combined to obtain a statistically significant average spectrum for the input galaxy sample.   This method has now been used to successfully obtain high redshift determinations for \ohi in both single-dish and interferometric observations \citep{lah2007,delhaize2013,rhee2013}. The technique has also been demonstrated for absorption line studies \citep[eg.][]{gereb2013,gereb2014}. Another method is intensity mapping, which rather than trying to stack the emission from individual sources and correct for any confusion effects that may arise, instead analyses the intensity field of emission as a whole, through auto-correlation or cross-correlation of the \hi intensity field with that derived from an optical spectroscopic catalogue.  In the last decade, great strides have been made in the application of this technique to measure the global \hi content of the Universe \citep{pen2009,chang2010}.  Lastly, independent optical redshifts can be used to measure the \hi properties of individual sources at a much lower significance level that would be possible from the blind \hi data alone, significantly increasing sample sizes for galaxy-by-galaxy analyses.

\vspace{-2mm}
\section{Multiwavelength Facilities \& Surveys}
\vspace{-2mm}
\label{sec:facilities}

\begin{figure}
\hspace{-1.0cm}\includegraphics[trim=0mm 0mm 0mm 0mm,width=9cm,keepaspectratio=true,clip]{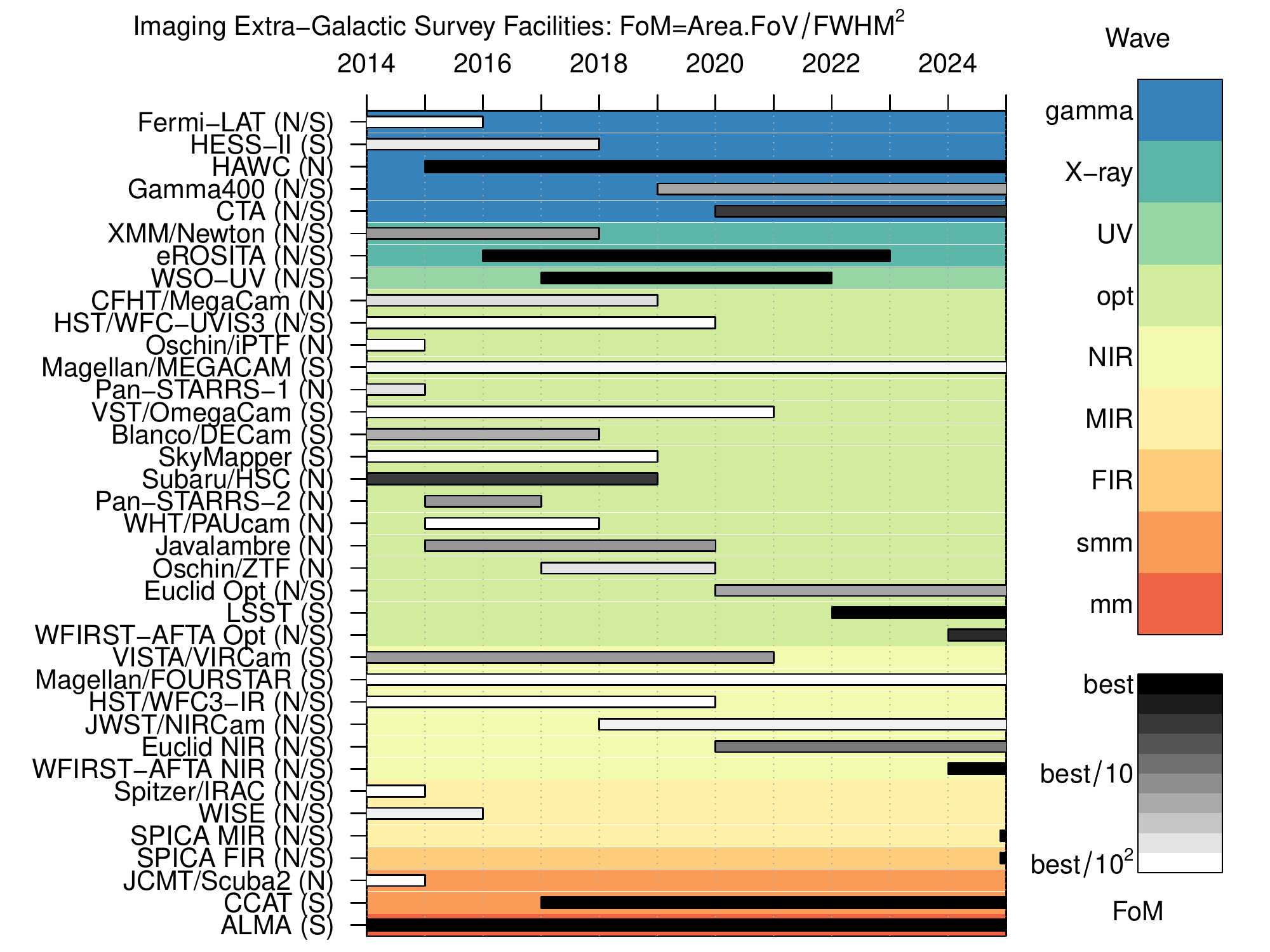}

\vspace{-6.5cm}

\hspace{8.0cm}\includegraphics[trim=0mm 0mm 0mm 0mm,width=9cm,keepaspectratio=true,clip]{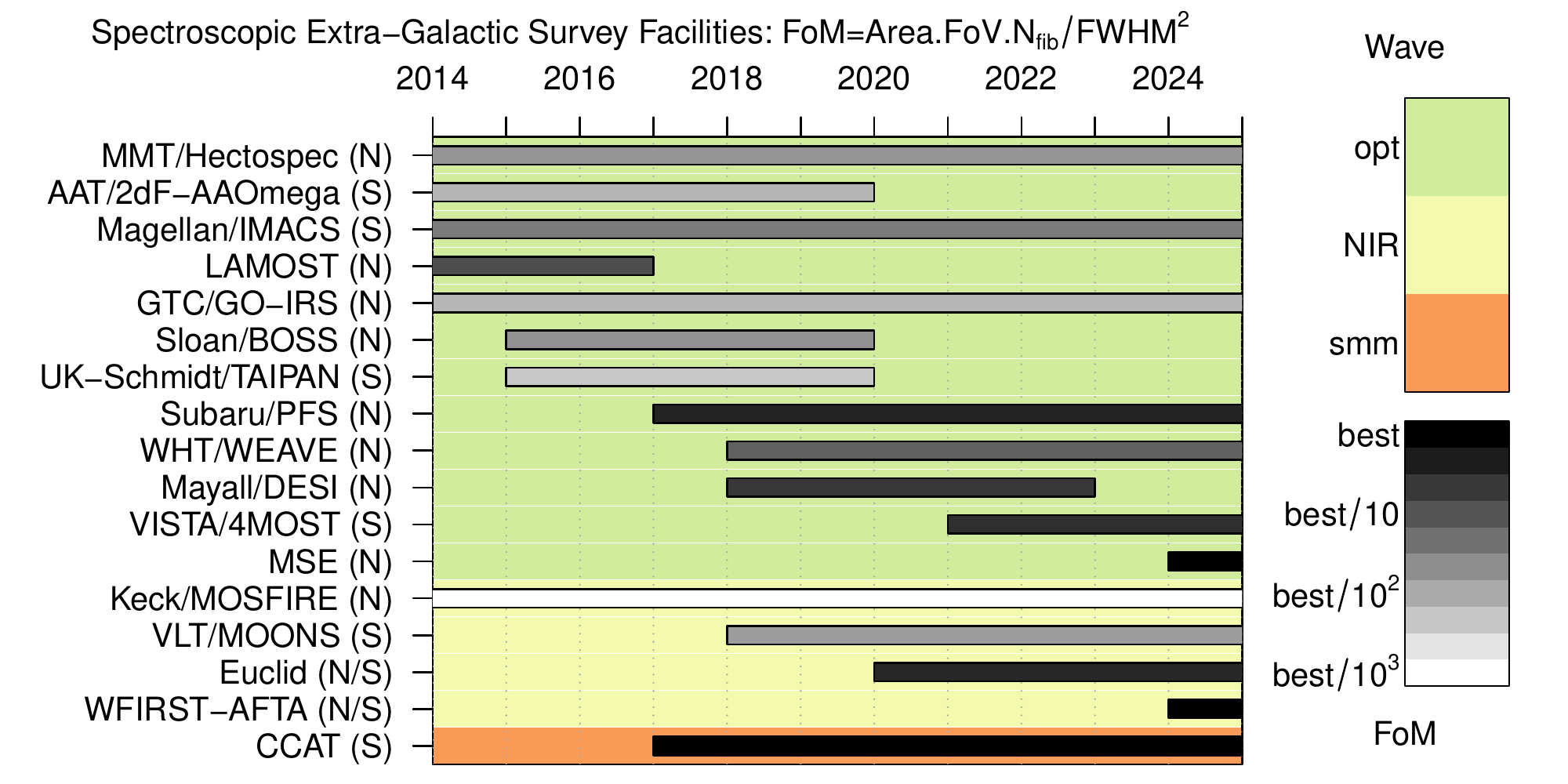}

\vspace{3.0cm}

\hspace{-1.0cm}\includegraphics[trim=0mm 0mm 0mm 0mm,width=9cm,keepaspectratio=true,clip]{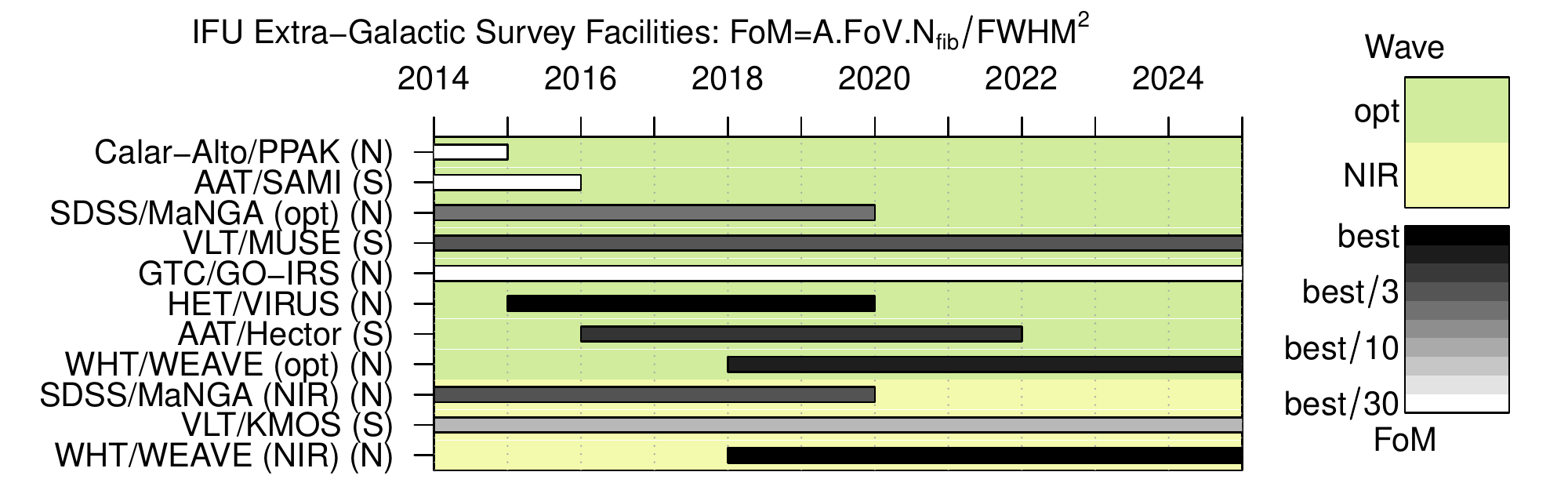}

\vspace{-5.0cm}

\hspace{8.0cm}\includegraphics[trim=0mm 0mm 0mm 0mm,width=9cm,keepaspectratio=true,clip]{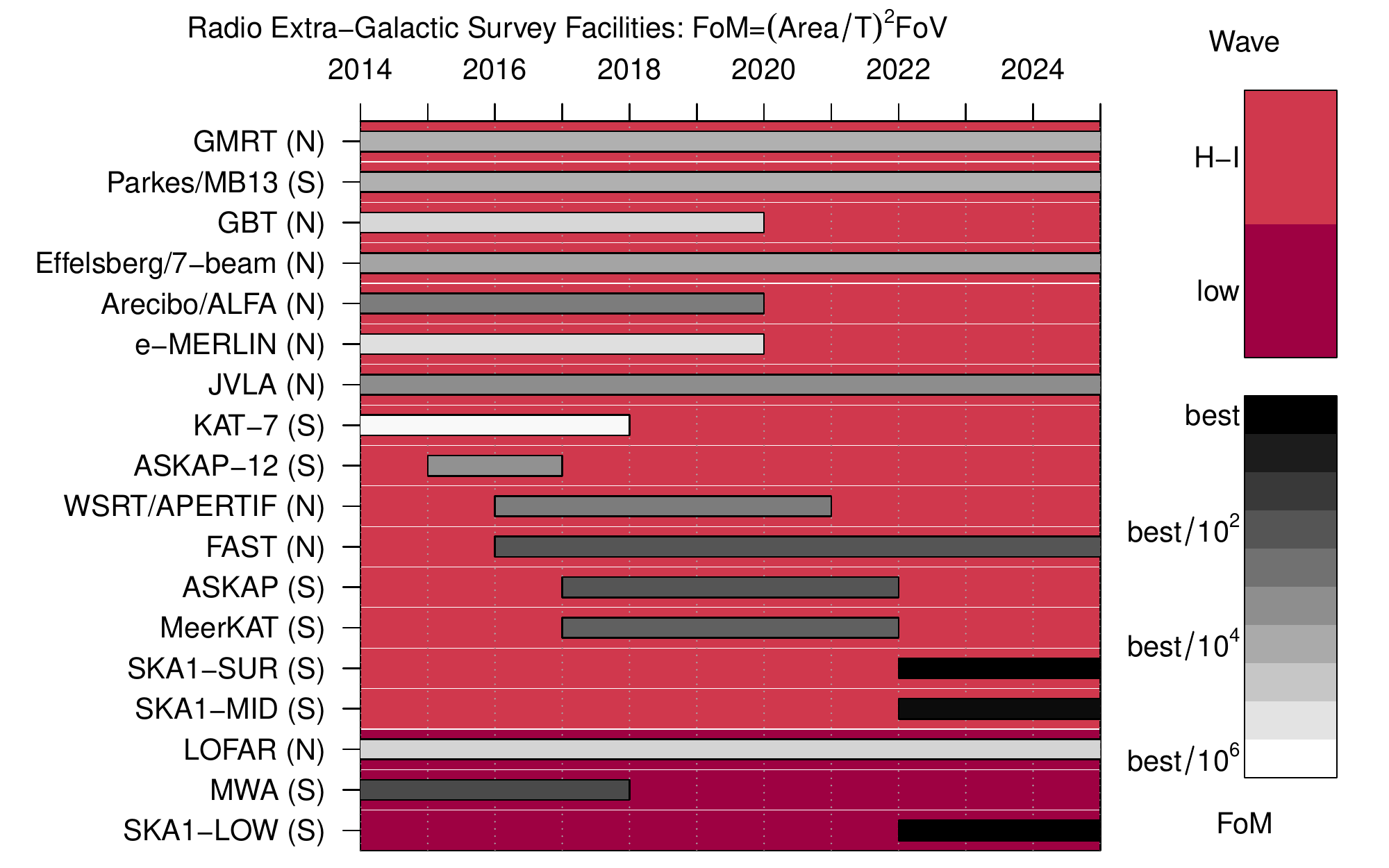}
\caption{
(top left) Comparison of various extra-galactic imaging survey facilities that will operate between now and 2025.  Figure of merit (FoM) is calculated using Area.FoV/(FWHM$^2$). (top-right) Comparison of  various extra-galactic spectroscopic survey facilities that will operate between now and 2025.  FoM is calculated using Area.FoV.N$_{fib}$/(FWHM$^2$). (bottom-left) Comparison of various extra-galactic IFU survey facilities that will operate between now and 2025. FoM is calculated using Area.FoV.N$_{fib}$/(FWHM$^2$). (bottom-right) Comparison of various extra-galactic radio survey facilities that will operate between now and 2025.  FoM is calculated using (Area/T)$^2$.FoV. `\hi' corresponds to facilities able to observe \hi in the local Universe, i.e. they can observe at frequencies as high as 1.4 GHz.  Note that in all panels FoM shading is scaled within a wavelength subset. The accompanying data for all charts is given in the Appendix.  Versions of these charts and table data can be generated online using an interactive tool located at https://asgr.shinyapps.io/ganttshiny.}
\label{fig:gantt}
\end{figure}

In this section, we review current and planned multiwavelength survey facilities, examining these by survey class (imaging, spectroscopic, IFU) and assessing their relative performance as a function of wavelength to identify those best suited to provide matching multiwavelength data for the strawman SKA \hi surveys.  Note that we have not differentiated between facilities with open data access policies and those with restricted proprietary periods.  However, over the lifetime of the SKA it is expected that all of these datasets will become publicly available, as is normal astronomy practice.

Given inevitable variations in survey design, there is not a single figure of merit (FoM) that covers all scientific purposes. For instance, for spectroscopic surveys an experiment designed to cover a huge area of sky at low redshift will need a large field of view but not necessarily huge multiplexing or collecting area given the distribution of sources on the sky. Equally, a high redshift study might need good multiplexing and a large collecting area but not a huge field of view. To allow for these design variations in desired FoMs we have designed an interactive website where users can build their own FoM, amongst other interactive options. The aim is to keep this database up-to-date with a moving 10-year window. The current host location is at: https://asgr.shinyapps.io/ganttshiny/.

\subsection{Optical/near-IR imaging surveys}

From a multiwavelength standpoint, imaging surveys represent the base from which other, more detailed follow-up observations can follow independently of the \hi selected dataset, and are therefore a crucial starting point. The survey imaging needs to be competitive otherwise all that follows will be similarly limited. 

There are a significant number of imaging facilities that have recently come online or will come online in the next decade. The top-left panel of Figure~\ref{fig:gantt} is a Gantt-Chart representation of when these facilities might become available, and how `fast' they are at surveying the extra-galactic sky. The chosen metric provides an insight into depth (telescope area) and sky coverage (FoV) per unit time, and data quality (FWHM image resolution). The majority of facilities will operate in the optical and there are two clear front-runners: Subaru Hyper-SuPrime-Cam, and towards the end of the decade, LSST. In the near-IR, arguably the more obvious band for sample selection as it most closely follows stellar mass, the obvious winner is VISTA, moving to Euclid towards the end of the decade.

The SKA will be largely accommodated by the huge Southern sky survey area and depth of LSST (at least in the optical). For this reason we have confidence that a large fraction of the imaging side of the SKA multiwavelength data will be either in place, or appearing, by the time SKA begins operations.  This is one part of the multiwavelength equation the SKA will not need to worry about going forward, but it should be noted that near-IR surveys will be less well matched in both depth and area.

\subsection{Massively multiplexed spectroscopic surveys}

Moving into the next generation of surveys, there will need to be a natural synergy with optical and NIR spectroscopic campaigns.  There are a number of multi-object spectroscopic (MOS) facilities available currently, and that are scheduled to come online between 2015--2025.  Figure~\ref{fig:gantt} (top-right) is a Gantt-Chart representation of when these facilities might become available, and how fast they are at spectroscopically surveying extra-galactic objects.

One of the fastest MOS facilities available is AAT/AAOmega. This has been used to conduct a number of multiple hundred thousand sized extra-galactic surveys: 2dFGRS, WiggleZ and GAMA.  Looking forward, there are a huge number of next generation MOS facilities scheduled to start operating in the Northern hemisphere before between 2017--2019, however only VISTA/4MOST will be operating in the South (i.e.\ overlapping substantially with ASKAP, MeerKAT and SKA).  For this reason alone it would be a compelling next generation MOS facility for the SKA to become involved in. It is worth noting that these next-generation MOS facilities are all comparably fast, and the reality is the faster extra-galactic surveys will be conducted on the most dedicated facilities.

\subsection{Spatially resolved spectroscopic surveys}

An equally important direction for next generation survey science is spatially resolved spectroscopy through the development of novel and pioneering instrumentation (e.g. AAT/SAMI, VLT/MUSE, VLT/KMOS).  A key power of these instruments is their ability to spatially resolve the chemistry and dynamics of galaxies -- star-formation, metallicity, stellar population ages, angular momenta and stellar dispersions.  Current spatially resolved surveys are at the thousand galaxy mark (the Spain-led CALIFA, the Australian-led SAMI and the US-led MANGA surveys).   ATLAS$^{\rm 3D}$ has also compiled a large sample of IFU and \hi measurements for early-type galaxies with WHT and WSRT.

The AAT/SAMI facility is the only multi-object IFU facility currently available, but others are scheduled to come online between 2015--2025.  Figure~\ref{fig:gantt} (bottom-left) is a Gantt-Chart representation of when these facilities might become available and how fast they are at surveying extra-galactic objects with IFUs.  Simultaneously probing gas and stellar kinematics through the combination of SKA and optical/NIR IFU data will be a huge science lever in the near future. Beyond this, high $S/N$ chemistry can provide great insight into the conversion of \hi and molecular gas into stars and metals, but a full understanding will require large numbers of galaxies in huge spatial and multiwavelength detail.

\vspace{-2mm}
\section{Optimal Matching of HI and Multiwavelength Datasets}
\vspace{-2mm}
\label{sec:design}

The minimal obvious requirement to maximise scientific return from SKA \hi surveys is to ensure that multiwavelength survey programs are carried out in the same regions of sky as those proposed for \hi. Beyond that, there are clearly pragmatic considerations regarding the depth and area of such overlapping multiwavelength surveys, with consideration needed for likely source densities, survey times and expense.  We perform a basic assessment of the optimal required depth for multiwavelength data as a function of SKA survey area and integration time by analysing simulated data from S3-SAX \citep{obreschkow2009x,2014arXiv1406.0966O}.  This data spans the redshift range $0<z<1.2$ in a 100 deg$^{2}$ volume-limited light-cone containing galaxy cold gas masses, \hi and H$_{2}$ masses, stellar masses, luminosities, and apparent magnitudes.  We apply apparent \hi, r-band and CO detection thresholds to assess the degree of overlap between a range of simulated observed galaxy datasets. For the detection of \hi sources by the SKA in this analysis, a 5$\sigma$ (optimal) threshold is applied.  A synthesised beam size of 5'' is used for all surveys, with the point source signal-to-noise value of galaxies being adjusted according to their size relative to the beam (SNR $\propto 1/\sqrt{1 + A_{\rm galaxy}/A_{\rm beam}}$).

\begin{figure}[t]
\footnotesize
\begin{center}
\includegraphics[trim=0mm 0mm 0mm 0mm,width=6.3cm,keepaspectratio=true,clip]{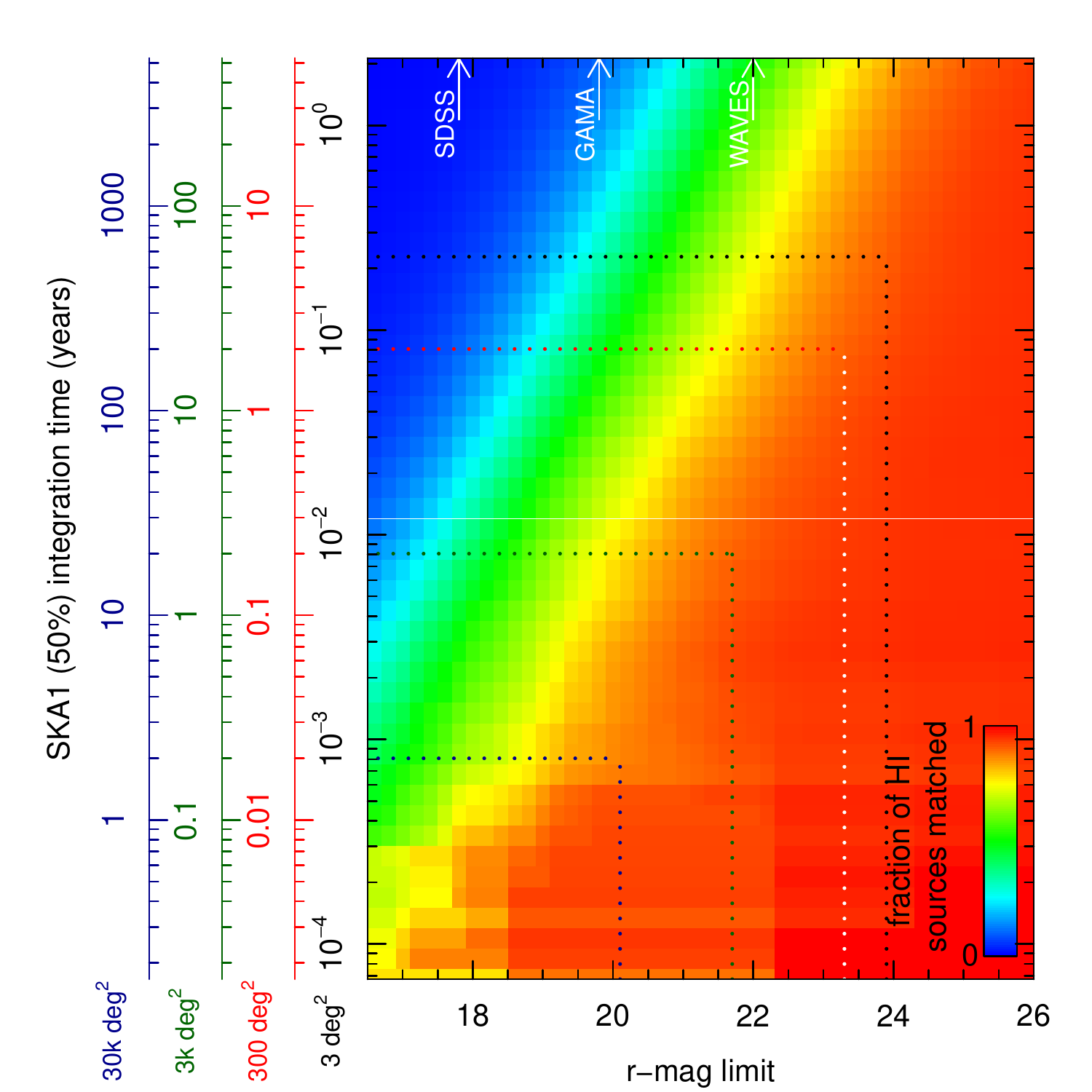}
\includegraphics[trim=0mm 0mm 0mm 0mm,width=6.3cm,keepaspectratio=true,clip]{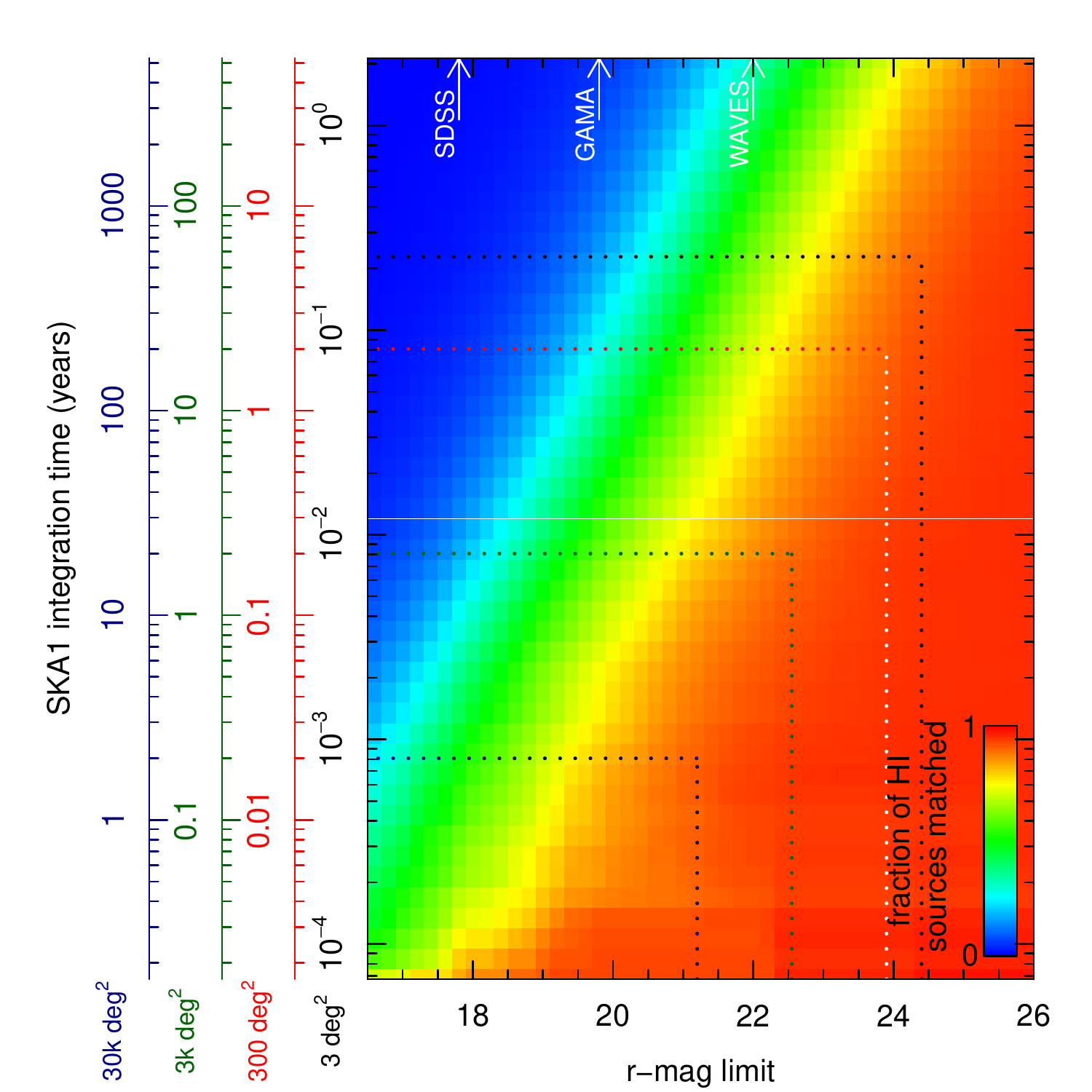}\\
\includegraphics[trim=0mm 0mm 0mm 0mm,width=6.3cm,keepaspectratio=true,clip]{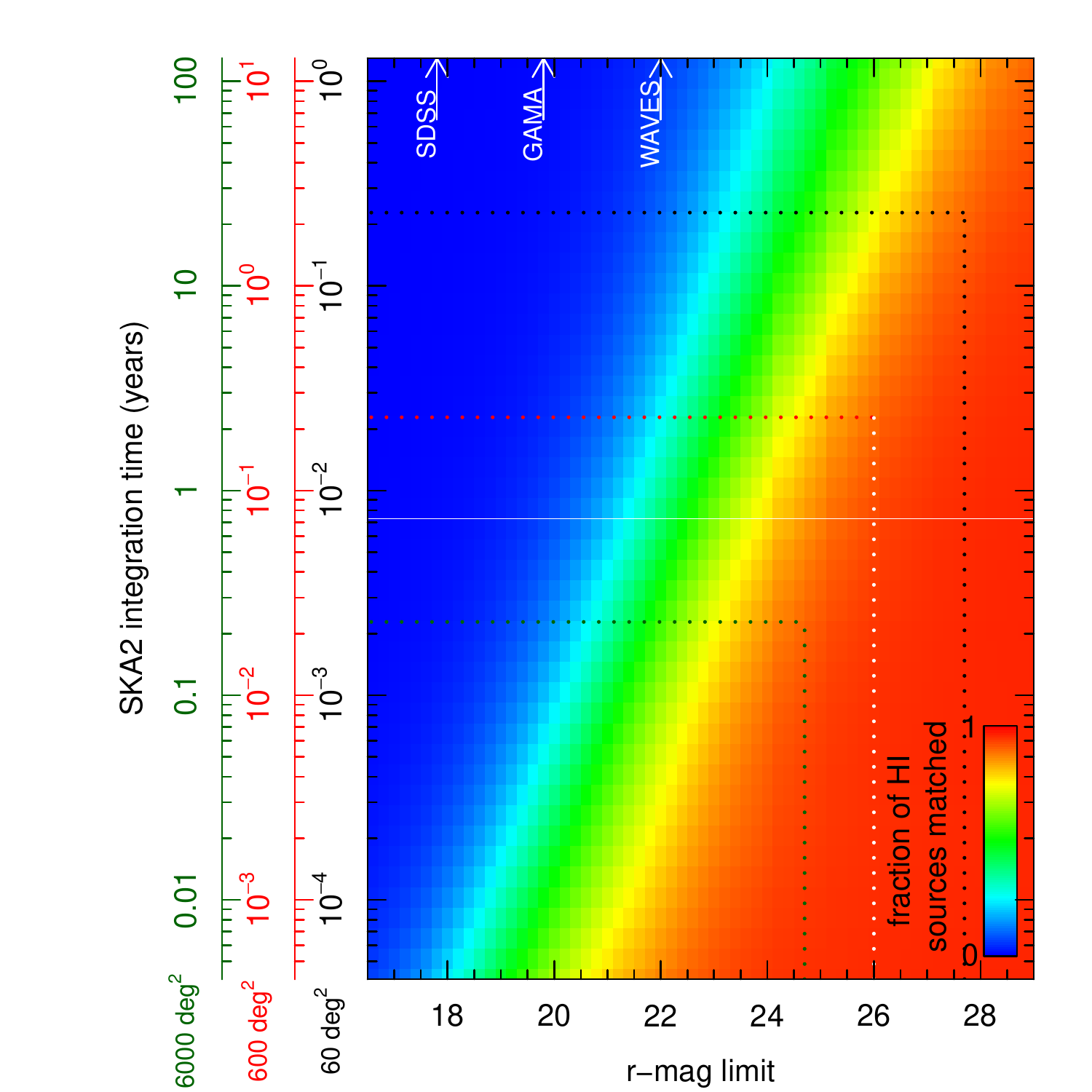}
\includegraphics[trim=0mm 0mm 0mm 0mm,width=6.3cm,keepaspectratio=true,clip]{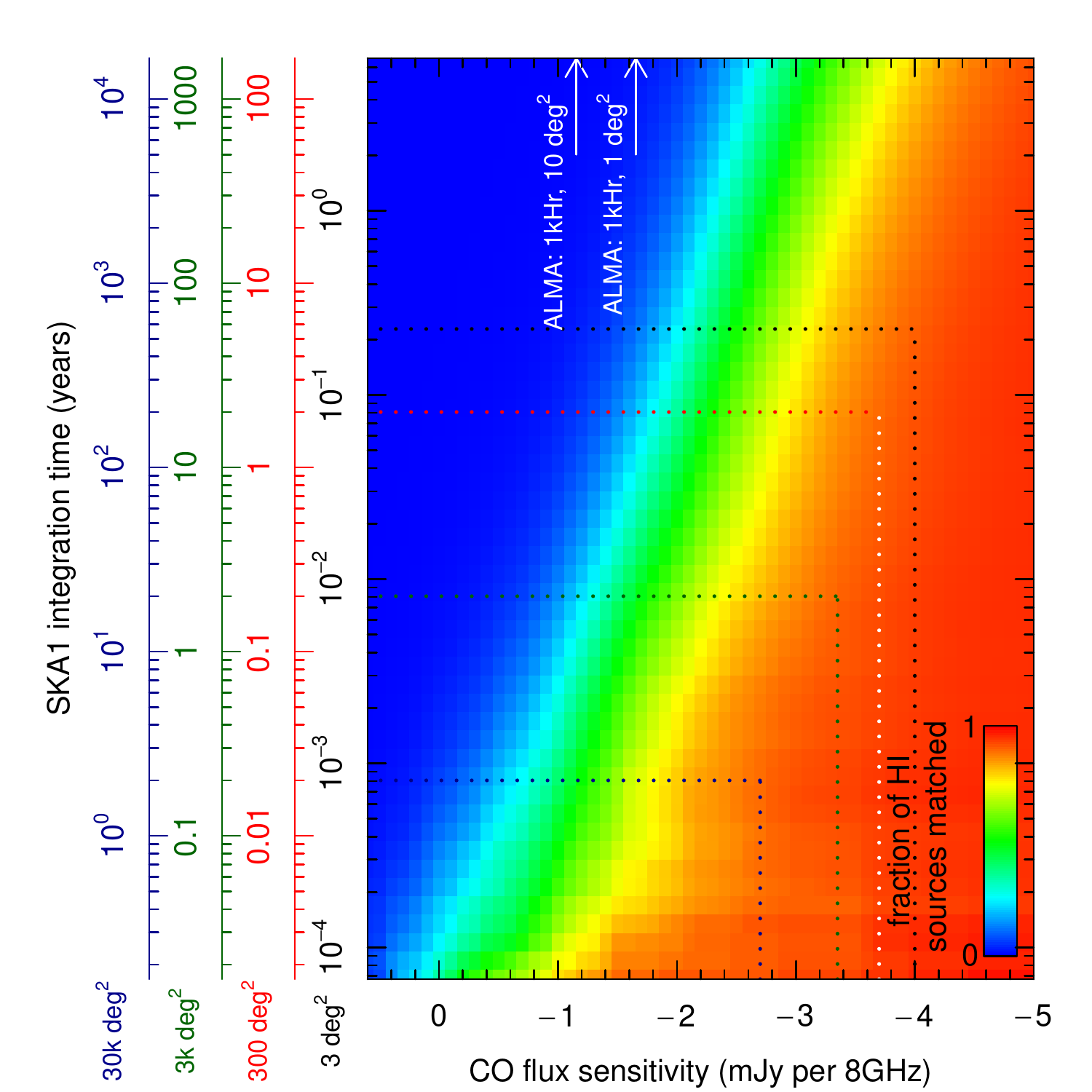}
\caption{(top-left, top-right, bottom-left) Fraction of \hi SKA survey sources (SKA1 50\%, SKA1, SKA2 respectively; as a function of integration time) that are detected in an r-band apparent magnitude-limited sample (down to a given magnitude).  The colour-scale indicates the value of the metric $F = \frac{A\cap B}{A}$. Different y-axis give the integration time scales for different survey areas.  For SKA1 50\% and SKA1, four surveys are considered: 3 deg$^{2}$ using SKA1-MID (and 50\%; black), 300 deg$^{2}$ using SKA1-SUR (and 50\%; red), 3000 deg$^{2}$ using SKA1-SUR (and 50\%; green), and 30000 deg$^{2}$ using SKA1-SUR (and 50\%; blue).  For SKA2, three surveys are considered: 60 deg$^{2}$ (black), 600 deg$^{2}$ (red), and 6000 deg$^{2}$ (green). Dotted lines indicate the r-magnitude limit needed to achieve matches for 90\% of sources in each SKA survey area (given 2000 hrs of integration for SKA1-MID \& SKA2 surveys, and 2 years of telescope time for SKA1-SUR surveys).  (bottom-right) Fraction of \hi SKA1 survey sources that are detected in a CO(1-0) flux sensitivity limited sample, with colour-scale, colour-axes and indication lines as before.}
\vspace{-6mm}
\label{fig:fij}
\end{center}
\end{figure}

\begin{figure}[t]
\footnotesize
\begin{center}
\includegraphics[trim=0mm 0mm 0mm 0mm,width=6.3cm,keepaspectratio=true,clip]{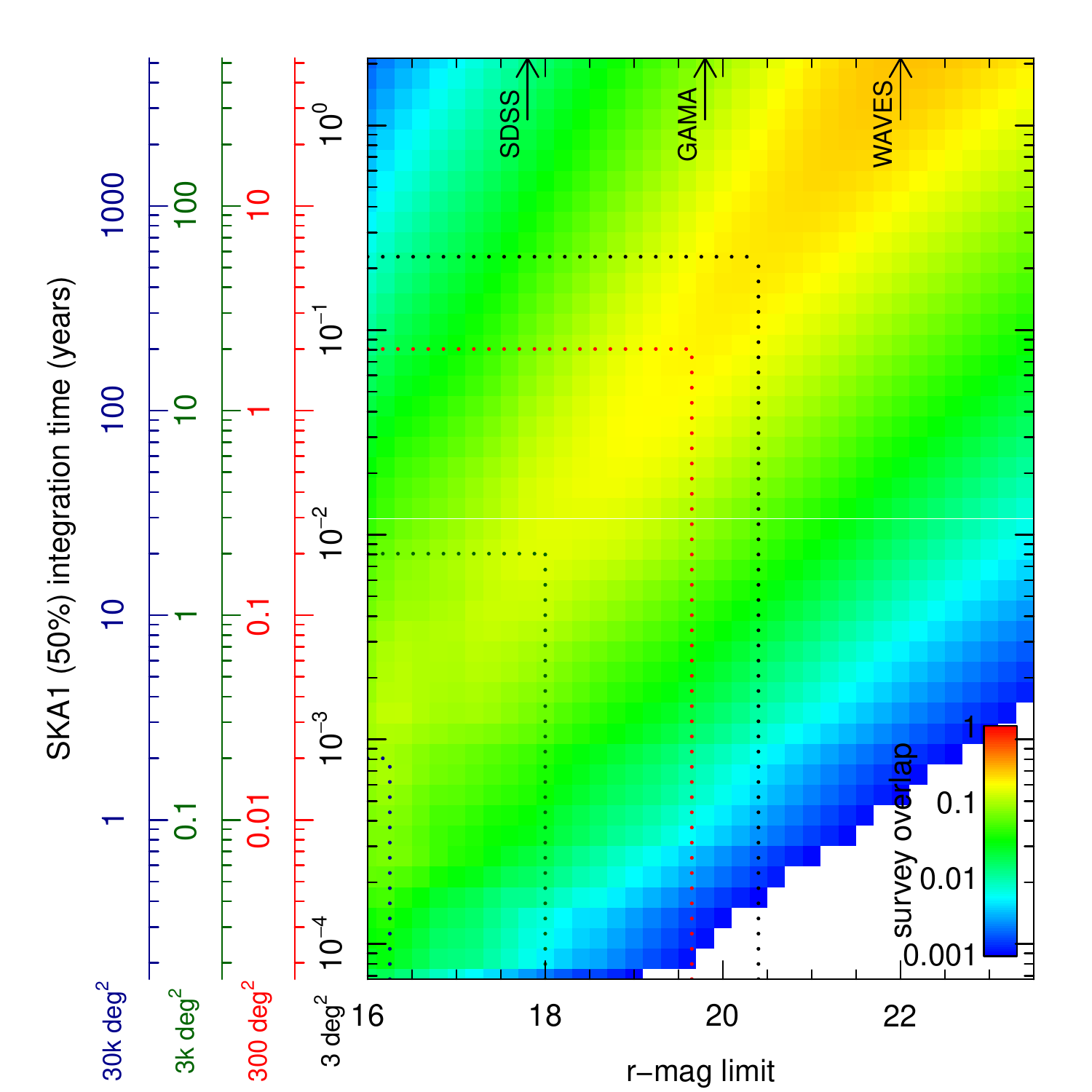}
\includegraphics[trim=0mm 0mm 0mm 0mm,width=6.3cm,keepaspectratio=true,clip]{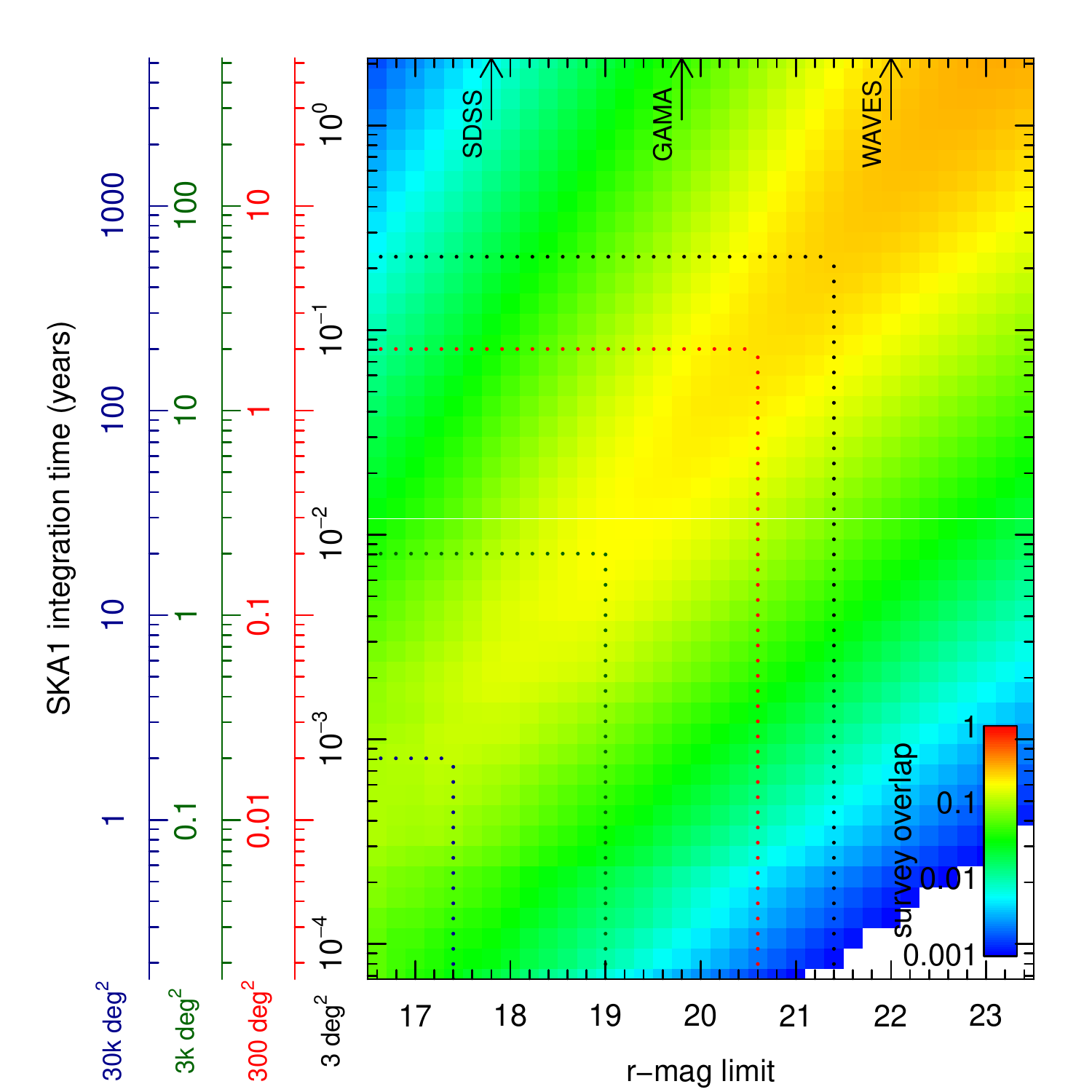}\\
\includegraphics[trim=0mm 0mm 0mm 0mm,width=6.3cm,keepaspectratio=true,clip]{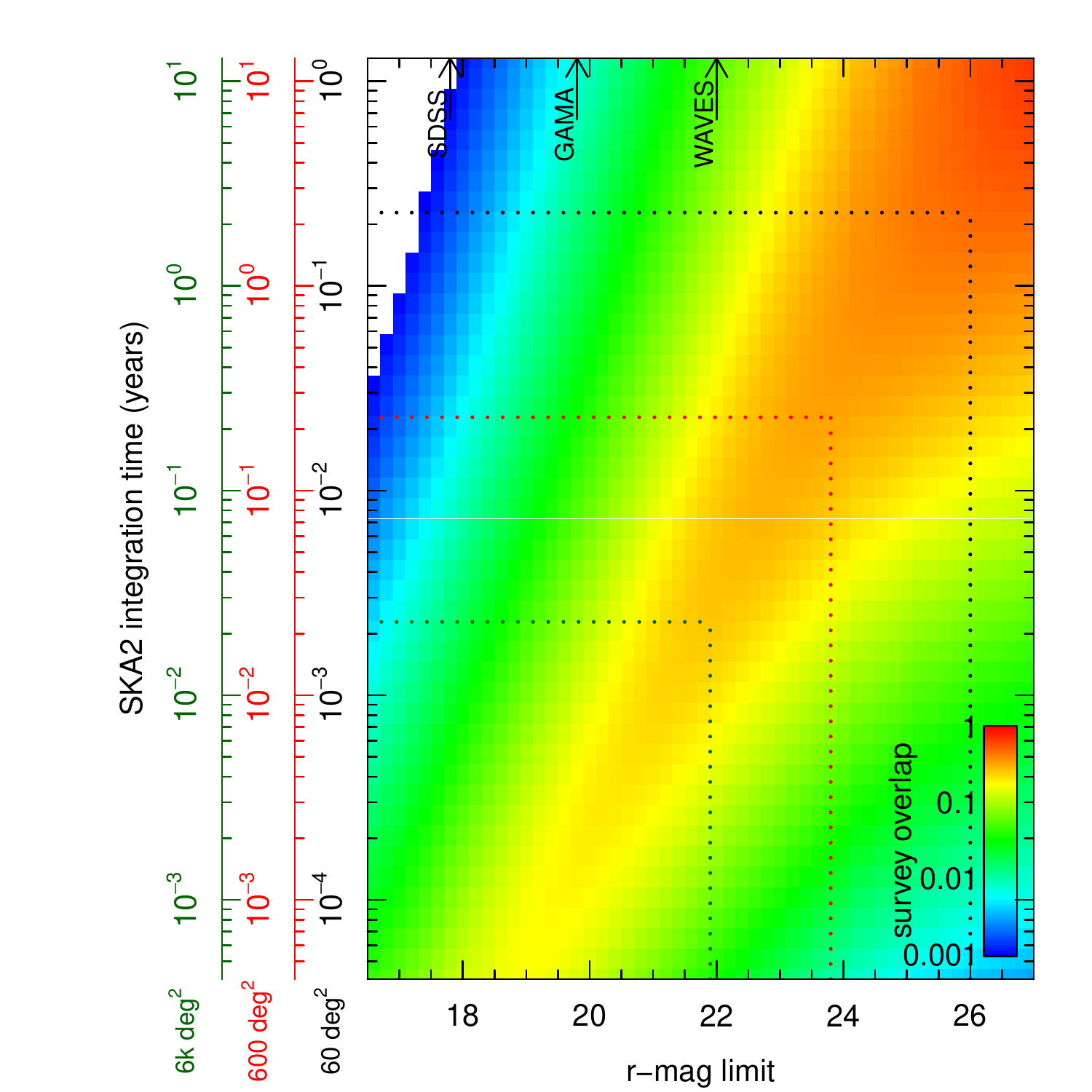}
\includegraphics[trim=0mm 0mm 0mm 0mm,width=6.3 cm,keepaspectratio=true,clip]{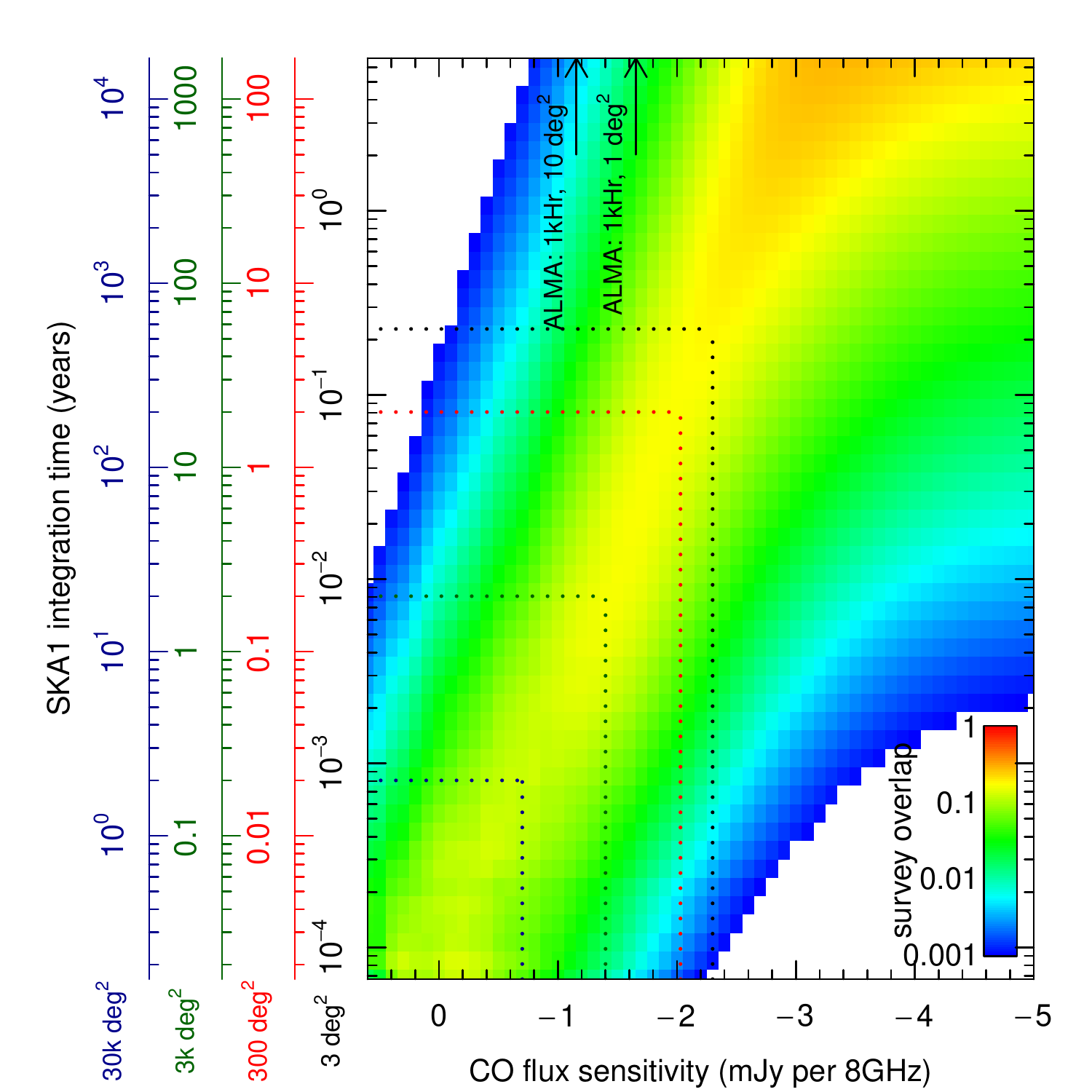}
\caption{(top-left, top-right, bottom-left) Degree of mutual overlap between the \hi sources detected by SKA telescopes (SKA1 50\%, SKA1, SKA2 respectively; as a function of integration time) and an optically selected r-band sample (down to a given magnitude).  The colour-scale indicates the value of the metric $E = \left(\frac{A\cap B}{A}\right)\left(\frac{A\cap B}{B}\right)$, where $E=0$ indicates that the two selected populations are disjoint, and $E=1$ indicates the selected populations are identical. Different y-axis give the integration time scales for different survey areas.  For SKA1 50\% and SKA1, four surveys are considered: 3 deg$^{2}$ using SKA1-MID (and 50\%; black), 300 deg$^{2}$ using SKA1-SUR (and 50\%; red), 3000 deg$^{2}$ using SKA1-SUR (and 50\%; green), and 30000 deg$^{2}$ using SKA1-SUR (and 50\%; blue).  For SKA2, three surveys are considered: 60 deg$^{2}$ (black), 600 deg$^{2}$ (red), and 6000 deg$^{2}$ (green). Dotted lines show the optimal r-magnitude matching limit for each SKA survey area (given 2000 hrs of integration for SKA1-MID \& SKA2 surveys, and 2 years of telescope time for SKA1-SUR surveys).  (bottom-right) Degree of mutual overlap between the \hi sources detected by SKA1 and a CO(1-0) flux sensitivity limited sample, with colour-scale, colour-axes and indication lines as before.}
\vspace{-6mm}
\label{fig:rbij}
\end{center}
\end{figure}

The first metric we consider is the simple fraction of \hi sources detected in a simulated \hi SKA survey (sample $A$) that will have a counterpart in an r-magnitude limited dataset (sample $B$): $F = \frac{A\cap B}{A}$.  r-band selection is used for redshift surveys such as GAMA due to the presence of H$\alpha$ (for higher redshift samples a longer observed frame wavelength selection may be preferable).  The results of this analysis are shown in Figure~\ref{fig:fij}, with separate panels given for SKA1 (50\%), SKA1 and SKA2.  We also include a plot for the match between SKA1 and a CO(1-0) flux limited dataset.  For SKA1 and its 50\% early deployment option, four separate y-axis are included for the four nominal reference surveys considered by the \hi Science Assessment Workshop (September 2013): 3 deg$^{2}$ (SKA1-MID and 50\% option), 300 deg$^{2}$ (SKA1-SUR and 50\% option), 3000 deg$^{2}$ (SKA1-SUR and 50\% option) and 30000 deg$^{2}$ (SKA1-SUR and 50\% option).  For SKA2, three survey areas are examined: 60 deg$^{2}$, 600 deg$^{2}$ and 6000 deg$^{2}$.  A horizontal dashed line indicates the reference integration times for each survey: 2000 hours for the 3 deg$^{2}$ survey with SKA1-MID and the 60-6000 deg$^{2}$ SKA2 surveys, and 2 years for the 300-30000 deg$^{2}$ surveys with SKA1-SUR.  The vertical intersect for each dashed survey line then corresponds to the required r-band limit for 90\% of the \hi sample to have an optical counterpart.

\begin{figure}[t]
\footnotesize
\begin{center}
\includegraphics[trim=0mm 48mm 0mm 5mm,width=7.5cm,keepaspectratio=true,clip]{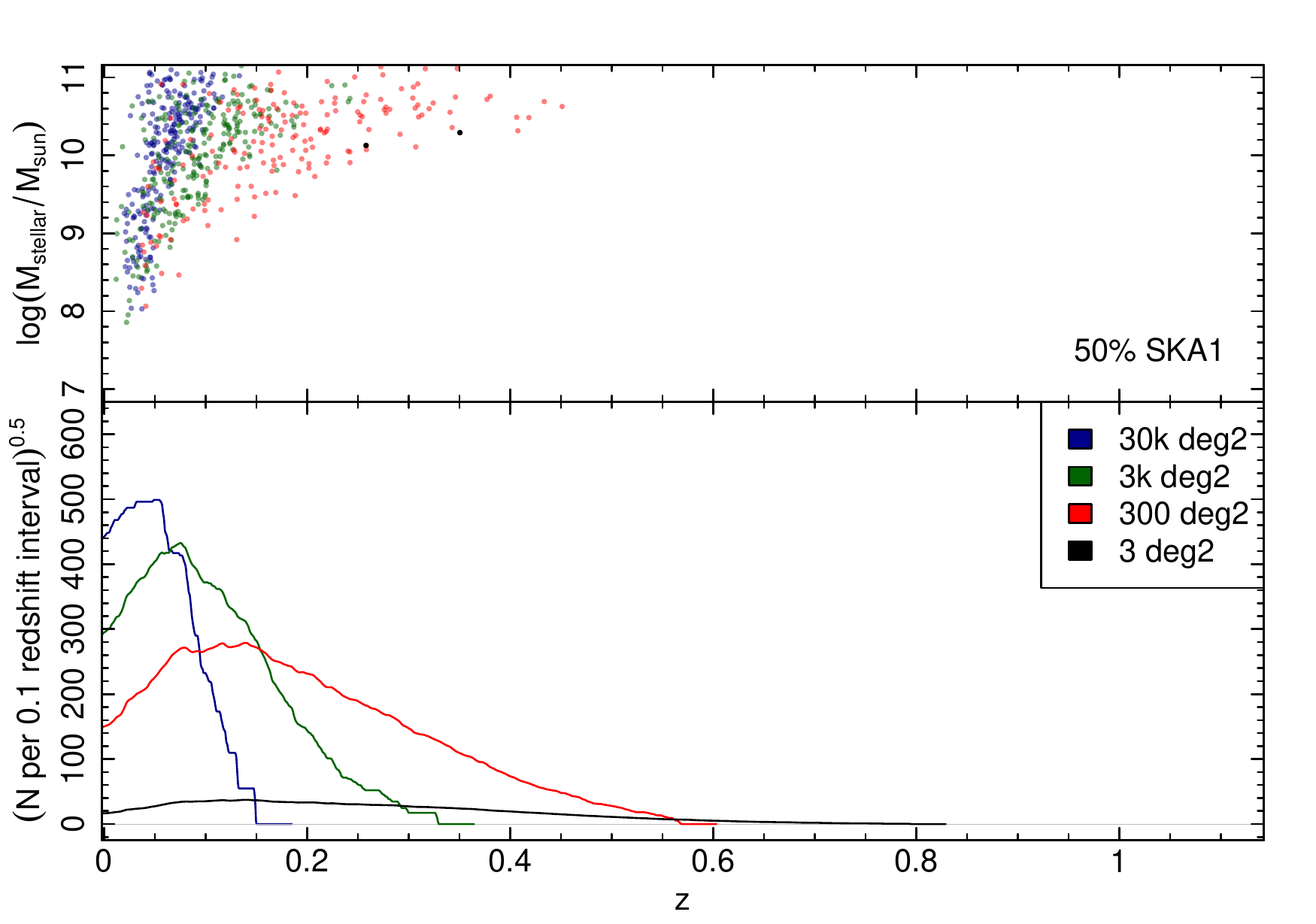}
\includegraphics[trim=0mm 48mm 0mm 5mm,width=7.5cm,keepaspectratio=true,clip]{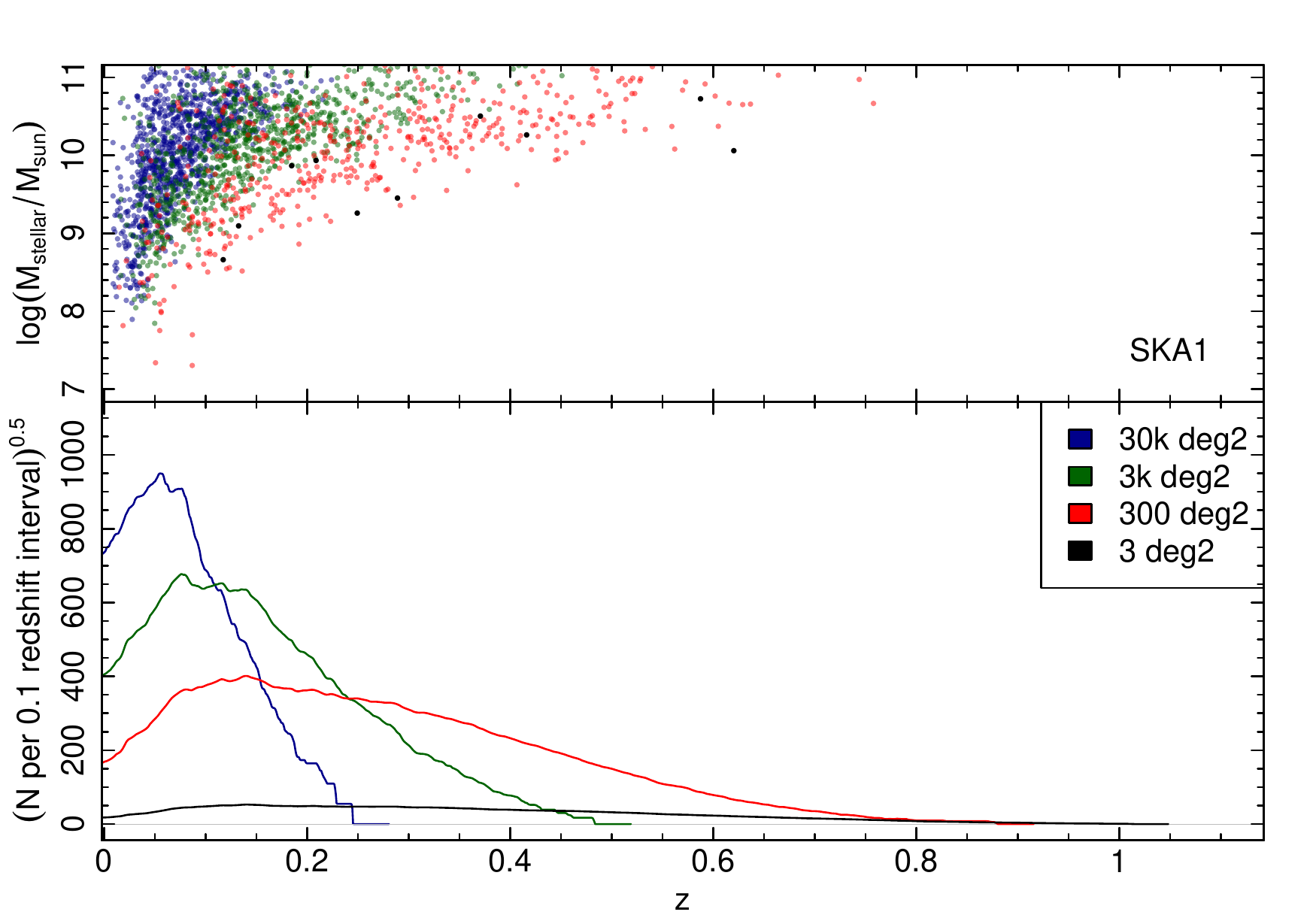}
\includegraphics[trim=0mm 48mm 0mm 5mm,width=7.5cm,keepaspectratio=true,clip]{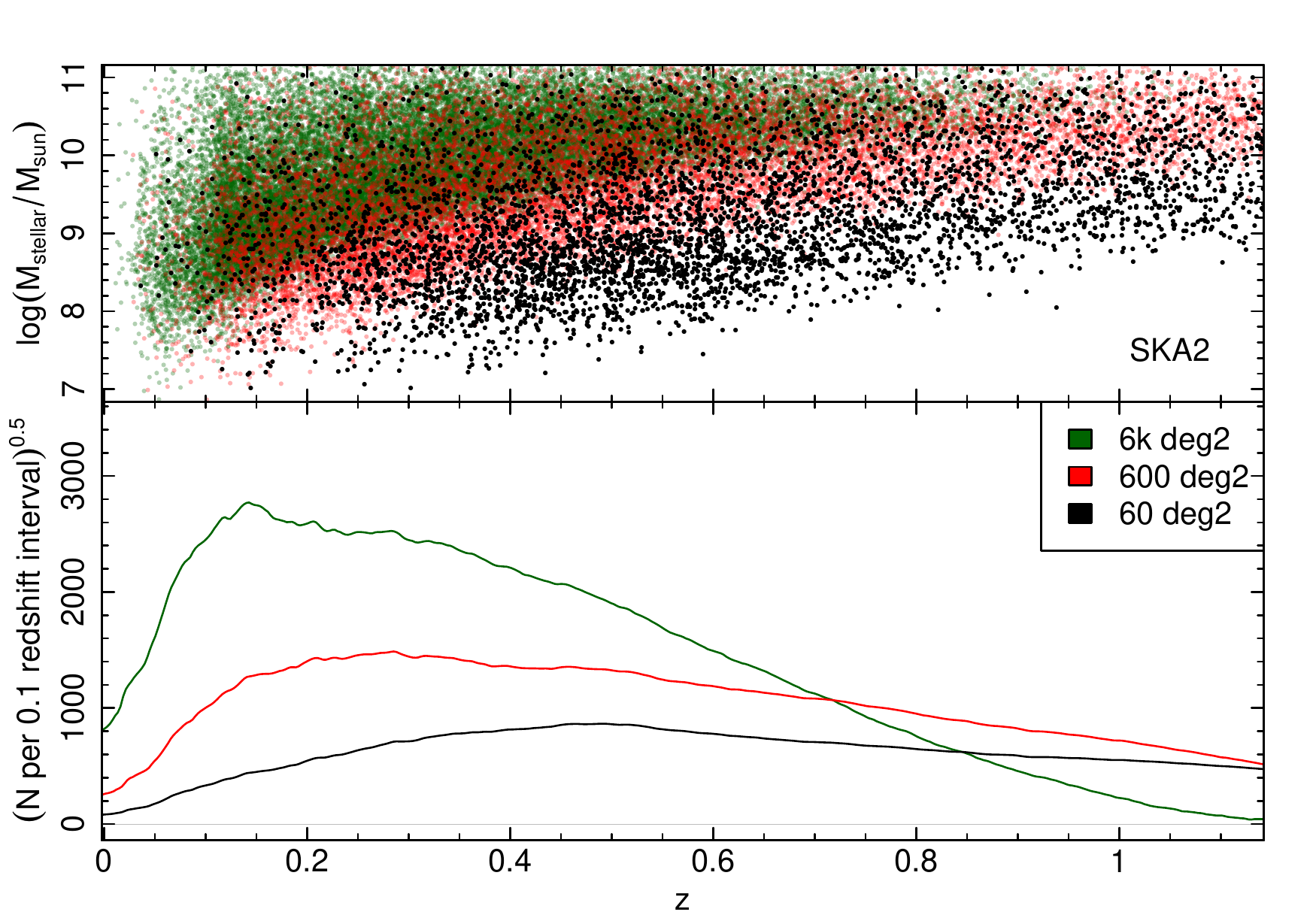}
\vspace{-1mm}
\caption{Stellar mass and redshift distributions of the optimal \hi and r-band limited samples as identified by the dashed lines in Figure~3 for SKA1 (50\%; top-left), SKA1 (top-right) and SKA2 (bottom).  Redshift distributions reflect the full number counts for each \hi and r-band limited survey, while one thousandth of each sample is plotted in the stellar mass distributions. \hi survey integration times are: 2000hrs (SKA1-MID, SKA1-MID 50\%: 3 deg$^{2}$; SKA2: 60, 600 and 6,000 deg$^{2}$); and 2 years (SKA1-SUR, SKA1-SUR 50\%: 300, 3000 and 30,000 deg$^{2}$).}
\label{fig:histograms}
\end{center}
\vspace{-6mm}
\end{figure}

While achieving high completeness (well-suited eg. for tasks such as optically motivated source finding, or an indicative limit for where \hi stacking gains interest), the above identified r-band samples will be comparatively inefficient when viewed from an optical standpoint, containing many sources that will have no \hi counterpart.  To assess the optimal r-band (and CO) depth at which the mutual overlap between \hi and multiwavelength samples is maximised, we consider a second metric $E = \left(\frac{A\cap B}{A}\right)\left(\frac{A\cap B}{B}\right)$, where a value of 0 indicates that the two selected populations are disjoint, and a value of 1 indicates the selected populations are identical.  The results of this analysis are shown in Figure~\ref{fig:rbij}, with panels arranged as for Figure~\ref{fig:fij}.  We note that the S3-SAX simulations used become incomplete for stellar/gas masses below $\sim10^{8}M_{\odot}$, a mass limit that may lead to an overestimate of survey overlap in the top-right portions of these diagrams compared to the full-mass distribution of galaxies in the Universe.  The dashed lines now indicate the r-band and CO flux thresholds corresponding to the optimal identified limits where the mutual overlap between the \hi and ancillary datasets are largest. The redshift and stellar mass distributions for the resultant \hi and r-band limited samples are shown in Figure~\ref{fig:histograms}.  

From these plots it can be seen that the optical imaging depth requirements for all the examined SKA surveys (up to, and including the full SKA2) will be well met by surveys such as those proposed for LSST.  Only the very deepest small-area \hi surveys may wish to coordinate with the LSST deep fields rather than the main survey (r $\sim$ 28 mag vs. $\sim$26.5 mag).  Black/white arrows are also included in the r-band plots to indicate the depths of some existing and planned low-SNR optical redshift surveys (SDSS, GAMA, WAVES).  These show that the optimal redshift sample depth requirements for SKA1-SUR and SKA1-SUR 50\% \hi surveys will be met by programs such as the WAVES design-reference survey on 4MOST (within its proposed $\lesssim 1000$ deg$^{2}$ area).  However, for SKA2, suitably matched redshift samples will require the development of a new class of multiplexed spectroscopic survey facility compared to the current generation of existing (and planned) 4m telescopes.  From the black arrows in the CO plots, showing example blind surveys that could be achieved with ALMA, it can be seen that the vast difference in survey speeds between ALMA and the SKA mean that ALMA observations will largely be restricted to targeting \hi sub-samples.
  
\vspace{-2mm}
\section{Conclusion}
\vspace{-1mm}
\label{sec:conclusion}

The SKA offers a unique opportunity to understand the evolution of galaxies through large-scale surveys of the most fundamental baryonic building block of the Universe: \hi.  The scientific opportunity provided by this capacity will be maximised by linking these data to that at other wavelengths, shedding light on the host of regulatory processes that govern the transfer and conversion of material from one baryonic state to another. At present, the planned capability for the acquisition of appropriate multiwavelength data for extragalactic \hi surveys varies significantly as a function of frequency and required survey area for SKA1.  Excellent optical/NIR imaging will be available through facilities such as LSST, Euclid and W-FIRST (indeed also extending to depths suitable for SKA2), caveat any issues that may arise through a need to coordinate target areas, or that potentially arise from restricted access to proprietary data.  However, there is a comparative lack of planned capacity to carry out complementary imaging surveys at other wavelengths.  
This will restrict the ability to trace important physical quantities such as star formation and dust content.  Molecular line observations are similarly constrained with only targeted follow-up observations currently possible over larger areas with facilities such as ALMA, rather than blind surveys of matching regions. The landscape for optical spectroscopy is significantly better, with suitable surveys expected to be possible with facilities such as VISTA/4MOST, Subaru/PFS, Mayall/DESI etc. However, the sensitivity of these facilities will likely only be good enough to provide redshift information, and not the required 10+ $S/N$ spectra required to trace chemistry which will need the development of 10+m dedicated spectroscopic survey facilities.  The availability of these various required multiwavelength data products can loosely be broken down into five categories as summarised below.\\

\vspace{-2mm}
\noindent Facility grades:
\vspace{-1mm}

\begin{enumerate}
\item Planned capacity exists, and blind ancillary survey data is expected to exist without coordination of target areas: optical imaging (LSST).
\item Planned capacity exists, and blind ancillary survey data could exist with coordination of target areas: NIR imaging (Euclid: < 8000 deg$^2$, W-FIRST); redshift emission spectra ($S/N\sim2$).
\item Planned capacity exists, and fully sampled ancillary data could exist, but only in a targeted follow-up mode of worthwhile sources: optical IFU; gas-phase emission spectra ($S/N\sim10$).
\item Planned capacity exists, and significantly sampled ($\sim1$--$10$\%) ancillary data could exist, but only in a targeted follow-up mode: mm (ALMA), FIR imaging (SPICA).
\item Little capacity is expected to be available in the coming decade, and only poorly sampled ($<1$\%) ancillary data are possible: UV imaging; stellar-phase absorption spectra ($S/N\sim100+$).
\end{enumerate}

\setlength\bibsep{0pt}
\footnotesize
\bibliographystyle{mn2e-v3}
\vspace{1mm}
\begin{multicols}{2}
\bibliography{mwave}
\end{multicols}

\newpage
\begin{sidewaystable}[ht]
\vspace{-10mm}
\section*{Appendix A: Multiwavelength Facilities}
\label{sec:appendix}

\noindent Data used for the Gantt charts presented in Figure~\ref{fig:gantt} are given below.  We note these numbers were compiled on a best-effort basis, with a degree of uncertainly or arbitrariness existing in some of the parameters (eg. dates). The exact specifications for a number of future facilities will no doubt also evolve with time.
\vspace{7mm}

\footnotesize
\centering
\begin{tabular}{lrrlllrllll}
  \hline
Facility &Start  &End    &Wave   & Hemi & \specialcell{Area\\ (m$^2$)} & \specialcell{A$_{\rm eff}$/T\\ (m$^2$/K)} & \specialcell{FoV\\(deg$^2$)} & \specialcell{FWHM\\(")} & Reference      \\  
  \hline
GMRT & 1995 & 20?? & \hi & N & 47,713 & 250 & 7.5E-02 & 2.0 & Ananthakrishnan S. \& Rao A.P., 2001, AP-RASC '01, 237 \\ 
  Parkes/MB13 & 1997 & 20?? & \hi & S & 3,217 & 100 & 4.8E-01 & 830 & Staveley-Smith L., et al., 1996, PASA, 13, 243 \\ 
  GBT & 2002 & 2020 & \hi & N & 7,854 & 276 & 1.5E-02 & 531 & Jewell P.R \& Prestage R.M., 2004, SPIE, 5489, 312 \\ 
  Effelsberg/7-beam & 2009 & 20?? & \hi & N & 7,854 & 262 & 1.1E-01 & 531 & Keller R., et al., 2006, Technical Report \\ 
  Arecibo/ALFA & 2010 & 2020 & \hi & N & 35,633 & 1150 & 2.4E-02 & 249 & Giovanelli, R., et al., 2005, AJ, 130, 2598 \\ 
  e-MERLIN & 2012 & 2020 & \hi & N & 9,073 & 60 & 2.4E-01 & 0.24 & Garrington S.T., et al., 2004, SPIE, 5489, 332 \\ 
  JVLA & 2012 & 20?? & \hi & N & 13,254 & 265 & 2.4E-01 & 1.5 & Perley R.A., et al., 2011, ApJL, 739, LL1 \\ 
  KAT-7 & 2012 & 2018 & \hi & S & 792 & 18 & 1.1E+00 & 287 & Booth R.S. \& Jonas J.L., 2012, AfrSk, 16, 101 \\ 
  ASKAP-12 & 2015 & 2017 & \hi & S & 1,357 & 22 & 3.0E+01 & 24 & Schinckel A.E., et al., 2012, SPIE, 8444 \\ 
  WSRT/APERTIF & 2016 & 2021 & \hi & N & 5,890 & 63 & 8.0E+00 & 19 & Oosterloo T., et al., 2009, PoS(SKADS 2009), 70 \\ 
  FAST & 2016 & 20?? & \hi & N & 70,686 & 2000 & 3.2E-02 & 177 & Nan R., et al., 2011, IJMPD, 20, 989 \\ 
  ASKAP & 2017 & 2022 & \hi & S & 4,072 & 65 & 3.0E+01 & 8.9 & Schinckel A.E., et al., 2012, SPIE, 8444 \\ 
  MeerKAT & 2017 & 2022 & \hi & S & 9,161 & 321 & 8.4E-01 & 6.6 & Booth R.S. \& Jonas J.L., 2012, AfrSk, 16, 101 \\ 
  SKA1-SUR & 2022 & 2027 & \hi & S & 14,674 & 391 & 1.8E+01 & 1.1 & Dewdney P., 2013, SKA-TEL-SKO-DD-001 \\ 
  SKA1-MID & 2022 & 2027 & \hi & S & 42,737 & 1630 & 6.7E-01 & 0.27 & Dewdney P., 2013, SKA-TEL-SKO-DD-001 \\ 
  LOFAR & 2012 & 20?? & low & N & 2,048 & 61 & 3.4E+00 & 0.33 & van Haarlem M.P., et al., 2013, A\&A, 556, AA2 \\ 
  MWA & 2013 & 2018 & low & S & 2,752 & 55 & 6.1E+02 & 120 & Tingay S.J., et al., 2013, PASA, 30, 7 \\ 
  SKA1-LOW & 2022 & 2027 & low & S & 876,485 & 1000 & 2.7E+01 & 11 & Dewdney P., 2013, SKA-TEL-SKO-DD-001 \\ 
\hline
\end{tabular}
\caption{Table comparing various extra-galactic radio survey facilities that will operate between now and 2025.  In the `Wave' column, `\hi' corresponds to facilities able to observe \hi in the local Universe, i.e.\ they can observe at frequencies as high as 1.4 GHz.  Most of the A$_{\rm eff}$/T values come from \citet{Dewdney:YebQLL23} which contains a good summary of radio facilities.  For a more detailed comparison of radio telescope performance characteristics beyond the first order approach used here, the reader is referred to \citet{Popping:2014wo} in this volume.}
\end{sidewaystable}

\begin{sidewaystable}[ht]
\footnotesize
\centering
\begin{tabular}{lrrllllllllrll}
  \hline
Facility &Start  &End    &Wave   & \specialcell{ $\lambda_{\rm low}$\\ (nm)}  & \specialcell{ $\lambda_{\rm hi}$\\ (nm)} & \specialcell{Spec.\\Res.} & Hemi & \specialcell{Area\\ (m$^2$)} & \specialcell{FoV\\(deg$^2$)} & \specialcell{FWHM\\(")} & N$_{\rm fibre}$ & Reference        \\
  \hline
\specialcell{MMT/\\Hectospec} & 2003 & 20?? & opt & 360 & 920 & 1,000 & N & 33.2 & 7.9E-01 & 0.7 & 300 & Fabricat D., et al., 2005, PASP, 117, 1411 \\ 
\specialcell{AAT/\\2dF-AAOmega} & 2006 & 20?? & opt & 370 & 950 & 1,300 & S & 11.9 & 3.1E+00 & 1.5 & 400 & Saunders W., et al., 2004, SPIE 5492 \\ 
  Magellan/IMACS & 2008 & 20?? & opt & 500 & 900 & 20 & S & 31.2 & 2.0E-01 & 0.7 & 2500 & Alison L. Coil, 2011, ApJ, 741, 8 \\ 
  LAMOST & 2012 & 2017 & opt & 370 & 900 & 1,000 & N & 12.6 & 2.0E+01 & 3.0 & 4000 & Wang X., et al., 2009, MNRAS, 394, 1775 \\ 
  GTC/GO-IRS & 2014 & 20?? & opt & 370 & 991 & 2,000 & N & 84.9 & 4.9E-02 & 0.8 & 1000 & Caballero J.A., 2011, HSA6 SEA, 769 \\ 
  Sloan/BOSS & 2015 & 2020 & opt & 360 & 1,000 & 1,560 & N & 4.9 & 7.1E+00 & 1.4 & 1000 & Gunn J.E., et al., 2006, AJ, 131, 2332 \\ 
\specialcell{UK-Schmidt/\\TAIPAN} & 2015 & 2020 & opt & 400 & 800 & 2,000 & S & 1.1 & 2.8E+01 & 1.5 & 300 &  Kuehn K., et al., 2014, SPIE, 9147\\ 
  Subaru/PFS & 2017 & 20?? & opt & 380 & 1,300 & 2,300 & N & 52.8 & 1.3E+00 & 0.7 & 2400 & Ian S., et al., 2012, SPIE 8446 \\ 
  WHT/WEAVE & 2018 & 20?? & opt & 370 & 1,000 & 5,000 & N & 13.9 & 3.1E+00 & 0.8 & 1000 & Balcells M., et al., 2010, SPIE, 7735 \\ 
  Mayall/DESI & 2018 & 2023 & opt & 360 & 980 & 2,000 & N & 12.6 & 7.1E+00 & 1.5 & 5000 & Flaugher B. \& Bebek C., 2014, SPIE, 9147 \\ 
  VISTA/4MOST & 2021 & 20?? & opt & 390 & 1,050 & 5,000 & S & 12.6 & 4.2E+00 & 0.7 & 2400 & de Jong R., et al., 2012, SPIE, 8446 \\ 
  MSE & 2024 & 2029 & opt & 370 & 1,300 & 2,000 & N & 78.5 & 1.8E+00 & 0.7 & 3200 & McConnachie A., et al., 2014, SPIE, 9145E \\ 
  Keck/MOSFIRE & 2012 & 20?? & NIR & 970 & 2,400 & 3,310 & N & 76.0 & 2.8E-03 & 0.7 &  46 & McLean I.S., et al., 2012, SPIE, 8446 \\ 
  VLT/MOONS & 2018 & 20?? & NIR & 800 & 1,800 & 4,000 & S & 50.3 & 1.3E-01 & 0.8 & 1000 & Cirasuolo M., et al., 2014, SPIE. 9147 \\ 
  Euclid & 2020 & 2026 & NIR & 1,100 & 2,000 & 250 & N/S & 1.1 & 5.0E-01 & 0.2 & 1700 & Beaulieu J.P., 2010, ASPC, 430, 266 \\ 
  WFIRST-AFTA & 2024 & 2030 & NIR & 1,100 & 2,000 & 200 & N/S & 2.4 & 2.8E-01 & 0.1 & 1000 & Green J., et al., 2012, arXiv, 1208.4012G \\ 
  CCAT & 2017 & 20?? & smm & 577,000 & 1,538,000 & 400 & S & 491.0 & 7.9E-01 & 2.5 & 100 & Woody D., et al., 2012, SPIE, 8444 \\ 
   \hline
\end{tabular}
\caption{Table comparing various extra-galactic spectroscopic survey facilities that will operate between now and 2025.}

\vspace{10mm}
\footnotesize
\centering
\begin{tabular}{lrrllllllllrll}
  \hline
Facility &Start  &End    &Wave   & \specialcell{ $\lambda_{\rm low}$\\ (nm)}  & \specialcell{ $\lambda_{\rm hi}$\\ (nm)} & \specialcell{Spec.\\Res.} & Hemi & \specialcell{Area\\ (m$^2$)} & \specialcell{FoV\\(deg$^2$)} & \specialcell{FWHM\\(")} & N$_{\rm fibre}$ &  Reference        \\
  \hline
\specialcell{Calar-Alto/\\PPAK} & 2010 & 2015 & opt & 370 & 700 & 850 & N & 9.6 & 2.8E-04 & 1.5 & 400 & S.F.Sanchez et al., 2012, A\&A, 538, 8 \\ 
  AAT/SAMI & 2013 & 2016 & opt & 370 & 950 & 1,300 & S & 11.9 & 7.9E-01 & 1.5 & 400 & Croom S.M., et al., 2012, MNRAS, 421, 872 \\ 
  SDSS/MaNGA (opt) & 2014 & 2020 & opt & 360 & 1,000 & 2,000 & N & 4.9 & 7.1E+00 & 1.5 & 1423 & Law D.R., et al., 2014, AAS, 22325431 \\ 
  VLT/MUSE & 2014 & 20?? & opt & 460 & 930 & 2,000 & S & 50.3 & 2.8E-04 & 0.2 & 90000 & Bacon R., et al., 2010, SPIE, 7735 \\ 
  GTC/GO-IRS & 2014 & 20?? & opt & 370 & 991 & 2,000 & N & 84.9 & 7.1E-04 & 0.8 & 1600 &  Caballero J.A., 2011, HSA6 SEA, 769  \\ 
  HET/VIRUS & 2015 & 2020 & opt & 350 & 550 & 550 & N & 64.2 & 1.1E-01 & 1.5 & 33600 & Hill G.J., et al., 2008, ASP conf, 399, 115 \\ 
  AAT/Hector & 2016 & 2022 & opt & 370 & 950 & 1,300 & S & 11.9 & 3.1E+00 & 1.5 & 3050 & Bland-Hawthorn J, 2014, arXiv, 1410.3838 \\ 
  WHT/WEAVE (opt) & 2018 & 20?? & opt & 370 & 1,000 & 5,000 & N & 13.9 & 3.1E+00 & 0.8 & 1000 & Balcells M., et al., 2010, SPIE, 7735 \\ 
  SDSS/MaNGA (NIR) & 2014 & 2020 & NIR & 360 & 1,000 & 2,000 & N & 4.9 & 7.1E+00 & 1.5 & 1423 & Law D.R., et al., 2014, AAS, 22325431 \\ 
  VLT/KMOS & 2014 & 20?? & NIR & 800 & 2,500 & 1,800 & S & 50.3 & 1.1E-02 & 0.7 & 4704 & Sharples R.M., et al., 2004, SPIE, 5492 \\ 
  WHT/WEAVE (NIR) & 2018 & 20?? & NIR & 370 & 1,000 & 5,000 & N & 13.9 & 3.1E+00 & 0.8 & 1000 & Balcells M., et al., 2010, SPIE, 7735 \\ 
   \hline
\end{tabular}
\caption{Table comparing various extra-galactic IFU survey facilities that will operate between now and 2025.}
\end{sidewaystable}

\begin{sidewaystable}[ht]
\footnotesize
\centering
\begin{tabular}{L{3cm}rrllllllL{6cm}}
  \hline
Facility &Start  &End    &Wave   & Hemi & \specialcell{Area\\ (m$^2$)} & \specialcell{FoV\\(deg$^2$)} & \specialcell{FWHM\\(")} & Reference      \\  
  \hline
Fermi-LAT & 2008 & 2016 & gamma & N/S & 2.5 & 2.8E+03 & 360 & Atwood W.B., et al., 2009, ApJ, 697, 1071 \\ 
  HESS-II & 2012 & 2018 & gamma & S & 50,000 & 8.0E+00 & 900 &  Becherini Y., et al., 2008, AIPC, 1085, 738\\ 
  HAWC & 2015 & 2030 & gamma & N & 10,000 & 2.5E+04 & 2,700 & Abeysekara A.U., et al., 2012, AsPaP, 35, 641 \\ 
  Gamma400 & 2019 & 20?? & gamma & N/S & 0.8 & 2.8E+03 & 36 & Galper A.M., 2013, AdSpR, 51, 297 \\ 
  CTA & 2020 & 2030 & gamma & N/S & 500,000 & 2.0E+01 & 900 & Gibney E, 2014, Nat, 508, 297 \\ 
  XMM/Newton & 2000 & 2018 & X-ray & N/S & 0.4 & 2.0E-01 & 50 & Kirsch M.G.F., et al., 2004, SPIE, 5488 \\ 
  eROSITA & 2016 & 2023 & X-ray & N/S & 0.4 & 7.9E-01 & 25 & Chon G. \& Bohringer H., 2013, AN, 334, 478 \\ 
  WSO-UV & 2017 & 2022 & UV & N/S & 2.3 & 2.0E-01 & 0.2 & Shustov B., et al., 2009, Ap\&SS, 320, 187 \\ 
  CFHT/MegaCam & 2003 & 2019 & opt & N & 10.2 & 9.5E-01 & 0.7 &  Aune S., et al., 2003, SPIE 4841 \\ 
  HST/WFC-UVIS3 & 2009 & 2020 & opt & N/S & 4.5 & 1.3E-04 & 0.1 & Windhorst R., et al., 2011, ApJS, 193, 27 \\ 
  Oschin/iPTF & 2009 & 2015 & opt & N & 1.1 & 7.1E+00 & 1.5 & Law N.M., et al., 2009, PASP, 121, 1395 \\ 
  Magellan/MEGACAM & 2009 & 20?? & opt & S & 31.2 & 2.0E-01 & 0.7 &  Szentgyorgyi A., et al., 2012, SPIE, 8446 \\ 
  Pan-STARRS-1 & 2010 & 2015 & opt & N & 2.5 & 7.1E+00 & 1.0 &  Morgan J.S., et al., 2014, SPIE, 914 \\ 
  VST/OmegaCam & 2011 & 2021 & opt & S & 5.3 & 7.9E-01 & 0.7 &  Iwert O., et al., 2006, SPIE 6276 \\ 
  Blanco/DECam & 2013 & 2018 & opt & S & 12.6 & 3.8E+00 & 1.0 & Hand E., 2012, Nat, 489, 190 \\ 
  SkyMapper & 2014 & 2019 & opt & S & 1.4 & 5.7E+00 & 1.6 & Keller S.C., et al., 2007, PASA, 24, 1 \\ 
  Subaru/HSC & 2014 & 2019 & opt & N & 52.8 & 1.8E+00 & 0.5 & Miyazaki S., et al., 2012, SPIE, 8446 \\ 
  Pan-STARRS-2 & 2015 & 2017 & opt & N & 5.1 & 1.4E+01 & 1.0 & Morgan J.S., et al., 2014, SPIE, 914 \\ 
  WHT/PAUcam & 2015 & 2018 & opt & N & 13.9 & 7.9E-01 & 1.2 & Madrid F., et al., 2010, SPIE, 7735 \\ 
  Javalambre & 2015 & 2020 & opt & N & 4.9 & 7.1E+00 & 0.7 & Abramo L., et al., 2012, MNRAS, 423, 4 \\ 
  Oschin/ZTF & 2017 & 2020 & opt & N & 1.1 & 3.6E+01 & 1.5 & Bellm E., 2014, htu conf, 27 \\ 
  Euclid Opt & 2020 & 2026 & opt & N/S & 1.1 & 5.0E-01 & 0.2 & Beaulieu J.P., 2010, ASPC, 430, 266 \\ 
  LSST & 2022 & 2032 & opt & S & 55.4 & 9.6E+00 & 0.7 & Krabbendam V.L., 2008, SPIE, 7012 \\ 
  WFIRST-AFTA Opt & 2024 & 2030 & opt & N/S & 4.5 & 2.8E-01 & 0.1 & Green J., et al., 2012, arXiv, 1208.4012G \\ 
  VISTA/VIRCam & 2009 & 2021 & NIR & S & 13.2 & 2.1E+00 & 0.6 & Dalton G.B., et al., 2006, SPIE 6269 \\ 
  Magellan/FOURSTAR & 2009 & 20?? & NIR & S & 31.2 & 7.9E-03 & 0.7 &  Persson S.E., 2008, SPIE, 7014\\ 
  HST/WFC3-IR & 2009 & 2020 & NIR & N/S & 4.5 & 1.3E-04 & 0.1 & Windhorst R., et al., 2011, ApJS, 193, 27 \\ 
  JWST/NIRCam & 2018 & 2028 & NIR & N/S & 33.2 & 7.9E-05 & 0.04 &  Beichman C.A., et al., 2012, SPIE, 8442\\ 
  Euclid NIR & 2020 & 2026 & NIR & N/S & 1.1 & 5.0E-01 & 0.2 & Beaulieu J.P., 2010, ASPC, 430, 266 \\ 
  WFIRST-AFTA NIR & 2024 & 2030 & NIR & N/S & 4.5 & 2.8E-01 & 0.1 & Green J., et al., 2012, arXiv, 1208.4012G \\ 
  Spitzer/IRAC & 2003 & 2015 & MIR & N/S & 0.6 & 7.5E-03 & 1.7 & Fazio, G.G., et al., 2004, ApJSS, 154, 10 \\ 
  WISE & 2010 & 2016 & MIR & N/S & 0.1 & 5.0E-01 & 5.0 & Wright E.L., et al., 2010, AJ, 140, 1868 \\ 
  SPICA MIR & 2025 & 20?? & MIR & N/S & 9.6 & 2.0E-01 & 1.0 & Nakagawa  T., et al., 2014, SPIE, 9143 \\ 
  SPICA FIR & 2025 & 20?? & FIR & N/S & 9.6 & 2.0E-01 & 5.0 & Nakagawa  T., et al., 2014, SPIE, 9143 \\ 
  JCMT/Scuba2 & 2011 & 2015 & smm & N & 176.7 & 1.8E-02 & 14 & Holland W.S., et al., 2013, MNRAS, 430, 2513 \\ 
  CCAT & 2017 & 20?? & smm & S & 491 & 7.9E-01 & 2.5 & Woody D., et al., 2012, SPIE, 8444 \\ 
  ALMA & 2013 & 20?? & mm & S & 6,569 & 2.8E-05 & 0.01 & Brown R.L., 2004, AdSpR, 34, 555 \\ 
   \hline
\end{tabular}
\caption{Table comparing various extra-galactic imaging survey facilities that will operate between now and 2025.}
\end{sidewaystable}

\end{document}